\documentclass[twocolumn,5p]{elsarticle}
\usepackage{hyperref}
\usepackage{float}
\usepackage{verbatim} 
\usepackage{pgf-pie}
\usepackage{amsmath,amssymb,amsfonts}
\usepackage{multicol}
\usepackage{multirow}
\usepackage{etoolbox}
\usepackage{amsmath}
\usepackage{mathtools}
\usepackage{hyperref}
\usepackage{array}
\usepackage{soul}
\usepackage{xcolor}
\definecolor{highlightyellow}{RGB}{255, 255, 153} 
\definecolor{highlightred}{RGB}{255, 102, 102} 
\definecolor{highlightgreen}{RGB}{102, 255, 102} 

\newcolumntype{M}[1]{>{\centering\arraybackslash}m{#1}}

\usepackage{booktabs}
\usepackage{subcaption}
\usepackage{graphicx}
\usepackage{float}
\floatstyle{plaintop}
\restylefloat{table}
\usepackage[export]{adjustbox}

\journal{}

\usepackage{filecontents}

\begin{document}
\begin{frontmatter}

\title{FusionLungNet: Multi-scale Fusion Convolution with Refinement Network for Lung CT Image Segmentation}

\author[label1]{Sadjad Rezvani}
\ead{sadjadrezvani@shahroodut.ac.ir}

\author[label1]{Mansoor Fateh\corref{cor1}}
\ead{Mansoor\_fateh@shahroodut.ac.ir}

\author[label1]{Yeganeh Jalali}
\ead{jalali.yeganeh@shahroodut.ac.ir}

\author[label2]{Amirreza Fateh\corref{cor1}}
\ead{amirreza\_fateh@comp.iust.ac.ir}

\cortext[cor1]{Corresponding authors.}
\address[label1]{Faculty of Computer Engineering, Shahrood University of Technology, Shahrood, Iran.}
\address[label2]{School of Computer Engineering, Iran
	University of Science and Technology
	(IUST), Tehran, Iran}

\begin{abstract}
Early detection of lung cancer is crucial as it increases the chances of successful treatment. Automatic lung image segmentation assists doctors in identifying diseases such as lung cancer, COVID-19, and respiratory disorders. However, lung segmentation is challenging due to overlapping features like vascular and bronchial structures, along with pixel-level fusion of brightness, color, and texture. New lung segmentation methods face difficulties in identifying long-range relationships between image components, reliance on convolution operations that may not capture all critical features, and the complex structures of the lungs. Furthermore, semantic gaps between feature maps can hinder the integration of relevant information, reducing model accuracy. Skip connections can also limit the decoder's access to complete information, resulting in partial information loss during encoding. To overcome these challenges, we propose a hybrid approach using the FusionLungNet network, which has a multi-level structure with key components, including the ResNet-50 encoder, Channel-wise Aggregation Attention (CAA) module, Multi-scale Feature Fusion (MFF) block, self refinement (SR) module, and multiple decoders. The refinement sub-network uses convolutional neural networks for image post-processing to improve quality. Our method employs a combination of loss functions, including SSIM, IOU, and focal loss, to optimize image reconstruction quality. We created and publicly released a new dataset for lung segmentation called LungSegDB, including 1800 CT images from the LIDC-IDRI dataset (dataset version 1) and 700 images from the Chest CT Cancer Images from Kaggle dataset (dataset version 2). Our method achieved an IOU score of 98.04, outperforming existing methods and demonstrating significant improvements in segmentation accuracy. Both the dataset and code are publicly available (\href{https://github.com/sadjadrz/Lung-segmentation-dataset}{Dataset link}, \href{https://github.com/sadjadrz/FusionLungNet}{Code link}).
\end{abstract}


\end{frontmatter}

\section{Introduction}\label{introduction}
Lung cancer stand as a significant health concern that is on the rise. Early detection of lung cancer is essential to increase the chance of treatment  \cite{minna2002focus}. To address this need, advancements in medical imaging have led to the development of critical techniques, including lung image segmentation. Lung image segmentation is essential for accurately detecting and outlining lung structures in medical images, such as computed tomography (CT) scans and X-rays. The method has been highly popular for research purposes over the years because of its potential in diagnosing various lung-related or other diseases, such as covid-19  \cite{udugama2020diagnosing}, pneumonia \cite{van2013diagnosing}, and lung cancer \cite{mclean2018diagnosing}.

In the process of lung segmentation, there are inherent challenges related to the characteristics of lung tissue. These include overlaps between tissue features such as blood vessels and bronchial structures with other elements like airways and the chest wall. This complexity makes precise delineation of tissue boundaries difficult, thereby reducing the accuracy and reliability of lung tissue segmentation \cite{mansoor2015segmentation}. During manual segmentation, radiologists rely on their experience to accurately outline tissue boundaries. However, this method has challenges such as depending heavily on individual skill, being time-consuming, having differences among operators, and the potential for human errors. These difficulties can affect the accuracy of segmentation and disease diagnosis, highlighting the importance of advanced automated methods and improvements \cite{patil2013medical}. Therefore, the exploration and advancement of automated segmentation techniques are imperative to overcome the limitations posed by manual segmentation and to ensure more accurate and efficient lung analysis in medical imaging.

In recent years, deep learning methods and their variants in medical imaging have got widespread attention  \cite{de2023deep,halder2023atrous,nitha2024novel}. The rise of deep convolutional neural networks (DCNNs) marked a notable advancement in overcoming the mentioned challenges \cite{yoo2015deep}. Following this, fully convolutional neural networks (FCNs) were suggested for the purpose of image segmentation \cite{long2015fully}. These networks accomplish pixel-level segmentation by reducing the resolution of input feature maps. Subsequently, the U-Net  \cite{ronneberger2015u} model was introduced as the foremost segmentation network among encoder-decoder models. U-Net is distinguished by its unique encoder and decoder layers, linked through skip connections. These connections effectively blend characteristics from both high-level and low-level image features, resulting in enhanced model accuracy and addressing the issue of vanishing gradients.

 Nowadays, researchers have proposed various versions of the U-Net network including NAUNet \cite{yang2023naunet}, IterMiUnet \cite{kumar2023itermiunet}, EU-Net \cite{yu2023eu}, MSIU-Net \cite{wu2023multi}, IBA-U-Net \cite{chen2021iba}, Wavelet U-Net++ \cite{agnes2024wavelet} and etc. 
While U-Net family networks have made substantial strides in medical segmentation, but they face several unique challenges. U-Net-based networks inherently struggle to capture long-range dependencies due to the local nature of convolutional operations. This can limit their ability to fully capture intricate details, especially in smaller structures. Additionally, the semantic gap between encoder and decoder feature maps, along with noise sensitivity, further impacts segmentation accuracy. Skip connections address this by transferring information from deeper network layers (encoding lower-resolution but larger contextual features) to shallower layers (encoding higher-resolution features). This helps the network consider both local details and contextual information. However, skip connections cannot fully recover lost or compressed information during the encoding process, leading to a semantic gap that may hinder the decoder from having all the necessary information for accurate segmentation \cite{chen2021iba,zhang2023sclmnet,wu2023deep}. These challenges collectively influence the practicality of these models in clinical applications.

Additionally, while some state-of-the-art models \cite{nitha2024novel,vijayakumar2025sustainable} utilize single loss functions to address segmentation tasks, they may not effectively capture the multifaceted nature of lung segmentation, leading to suboptimal performance. These challenges collectively influence the practicality of these models in clinical applications.

Given the mentioned challenges in lung segmentation, we have introduced an approach for lung CT segmentation known as FusionLungNet. FusionLungNet is an end-to-end, multi-level model that comprises two sub-networks: segmentation and refinement. The segmentation sub-network itself consists of several key components, including an encoder based on ResNet-50, a Channel-wise Aggregation Attention (CAA) module, a Multi-scale Feature Fusion (MFF) block, a self refinement (SR) module, and a decoder. First, the ResNet-50 encoder extracts important information from input images. These features are then structured into four blocks, with those beyond Block 4 forwarded to the CAA module. Here, the CAA module optimizes feature representation while mitigating computational complexity, ensuring the fidelity of target feature extraction. Following CAA, the MFF block operates strategically across decoder layers, fostering coherence between samples and merging information across diverse feature levels. Subsequently, the SR module resolves challenges related to multi-scale feature fusion modules by optimally transferring features to the decoder. This module can be used to correct minor defects in the original image. It can also be utilized to enhance the clarity and contrast of the original image. Thus, the image is processed by the self-correction module to produce a higher quality initial reconstruction image. During the training process, a hybrid loss function supervises the training stages to ensure accurate predictions. Our network's hybrid loss function consists of three metrics: Structural Similarity Index (SSIM) \cite{bakurov2022structural}, Intersection over Union (IOU) \cite{zhou2019iou}, and focal loss \cite{lin2017focal}. The combination of these metrics improves image reconstruction quality through several mechanisms. SSIM evaluates the perceived quality by comparing structural information between the original and reconstructed images, while IOU measures the overlap accuracy of predicted segments with the ground truth, ensuring spatial precision, and focal loss focuses on hard-to-classify pixels, reducing the impact of class imbalance. This comprehensive evaluation ensures that the reconstructed image retains structural integrity, spatial accuracy, and balanced classification. Using multiple metrics helps identify and correct different types of errors, enhancing detail preservation and making the model more robust to variations and noise. This balanced focus ensures that the reconstructed image is not only accurate in specific areas but also maintains high quality overall, leading to detailed, accurate, and robust results critical for medical imaging applications. 
 
The refinement sub-network, embedded within the residual enhancement module, further improves the quality of image reconstruction. Leveraging convolutional neural networks, this module conducts image reconstruction post-processing. It adeptly rectifies minor imperfections and enhancing clarity and contrast in reconstructed images. By seamlessly integrating both segmentation and refinement sub-networks, FusionLungNet presents a holistic solution to the challenges of lung CT segmentation, promising substantial advancements in image reconstruction quality and segmentation accuracy.
 
 The contributions of this paper can be summarized as follows. 
 \begin{itemize}
 	
 	\item \textbf{Introducing a novel dataset:} One of the main innovation of our work is the development and release of new labels for an existing lung CT image datasetS. We introduce LungSegDB, which features precise labeling of lung CT images from the LIDC-IDRI dataset (Version 1) and the Chest CT Cancer Images Dataset (Version 2). This is the first time these labels have been created, and we are publicly sharing them along with the images.
 	
 	\item \textbf{Channel-wise Aggregation Attention (CAA):} Our proposed approach introduces the CAA module, which enhances feature representation by emphasizing critical channels in feature maps. This allows the model to capture important details in lung CT images more effectively, enabling more accurate segmentation and reconstruction.
 	
 	\item \textbf{Multi-scale Feature Fusion module (MFF):} A noteworthy advancement in our approach involves enhancing the Feature Fusion Module. This module significantly contributes to elevating the quality of the reconstructed image by effectively combining features extracted from diverse sources.
 	\item \textbf{Hybrid loss function:}  Our network introduces a Hybrid Loss Function, incorporating three essential metrics including SSIM, IOU and Focal. This innovative combination of metrics is carefully designed to optimize the reconstruction quality.
 	
 	\item \textbf{Refinement sub-network:} Another distinctive feature is the refinement network \cite{rezvaniabanet}, which acts as an additional network for further refinement of the reconstructed image. 
 \end{itemize}

	The remainder of this paper follows a structured outline. Section \ref{sec:RelatedWork} provides an overview of related works in medical image segmentation, while section \ref{sec:proposedDataset} provides a description of the dataset and explain the labeling process. Section \ref{sec_Methodology} details the proposed method. Subsequently, Section \ref{sec_Experiments} presents comprehensive experiments and visualization analyses. The conclusion and future works for this work are summarized in Section \ref{sec_conclusion}.  
   
\section{Related work}\label{sec:RelatedWork}
	 In general, medical image segmentation methods can be categorized into two main groups \cite{saber2024efficient}: traditional methods and deep learning-based methods. Essentially, traditional methods rely on low-level visual features of images, while deep learning-based methods use deep neural networks to extract more intricate features. Next, we conduct an analysis of these two categories, highlighting key approaches within traditional methods and exploring the diverse landscape of deep learning-based methods. Also, we provide illustrative examples from each category to better elucidate their respective strengths and limitations.
\subsection{Traditional segmentation methods}\label{sec:RelatedWork.traditional}
     Traditional techniques for lung CT image segmentation encompass several methods. One of these methods is threshold-based segmentation, where different components of the image are identified by setting a threshold based on pixel intensities \cite{senthilkumaran2016image}. This approach is straightforward but may not be robust in cases of varying intensities or complex structures. Another common technique is the region-growing method, which detects homogeneous regions within the image \cite{mancas2005segmentation}. This method starts with a seed pixel and progressively expands regions with similar intensities, but it may struggle with noise or irregular shapes.  Watershed segmentation is another approach, employing watershed basins to identify sub-threshold points and transform them into basins \cite{hamarneh2009watershed}. While effective in certain scenarios, it can be sensitive to noise and may over-segment the image. Active contour models, including snake models, are also used \cite{chen2019learning}. These models combine image information and specific energies to automatically detect and move contours \cite{radeva1995snake}. While they are suitable for analyzing images with complex structures and lower levels of noise, they may struggle with highly variable organ shapes and contours.

     Despite their utility, traditional medical image segmentation methods face challenges with substantial variations in organ shapes and contours \cite{blaschke2004image}. This often results in segmentation results that fluctuate or decline in accuracy, especially as the complexity of segmented objects increases in practical applications. Consequently, ensuring satisfactory accuracy becomes increasingly difficult. The limitations of traditional medical image segmentation methods underscore the need for automated techniques with reliable algorithms, particularly in computer-aided diagnosis (CAD) applications for lung CT segmentation. Developing such techniques is crucial for enhancing segmentation accuracy and improving the reliability of diagnostic processes in medical imaging.

Techniques based on deep learning in the segmentation of lung CT images have been highly considered due to their superior ability to learn complex features as well as their ability to interact with big data [30]. One of the popular models in medical image segmentation, including lung CT images, is the U-Net [10] model. This model uses a convolutional neural network structure with encoder and decoder layers, which is specifically designed for segmentation applications. So far, many researchers have attempted to segment the lung with U-Net.  

\subsection{Deep learning-based segmentation methods} \label{sec:RelatedWork.deep}
Techniques based on deep learning in the segmentation of lung CT images have been highly considered due to their superior ability to learn complex features as well as their ability to interact with big data \cite{liu2021review}. One of the popular models in medical image segmentation, including lung CT images, is the U-Net \cite{russakovsky2015imagenet} model. This model uses a convolutional neural network structure with encoder and decoder layers, which is specifically designed for segmentation applications. So far, many researchers have attempted to segment the lung with U-Net. 

\subsubsection{U-Net based Methods} \label{sec:RelatedWork.deep.Unet}
   The good performance of U-Net has led to the development of various derivative architectures. For example, Liu et al. introduces a new method called MD-UNet for accurately extracting lesion areas in medical images  \cite{liu2024md}. They utilized some modules like Mixed depth-wise convolution attention block (MDAB) and Mixed depth-wise convolution residual module (MDRM) to mitigate intra-class differences and enhance segmentation accuracy. Kumar et al. introduced IterMiUnet, a lightweight convolution-based segmentation model for automatic blood vessel segmentation in fundus images \cite{kumar2023itermiunet}. IterMiUnet reduces parameters significantly while maintaining performance comparable to existing models by combining the segmentation capabilities of Iternet with the encoder-decoder structure of MiUnet. However, this may lead to a slight decrease in accuracy and other evaluation metrics. Additionally, this method may limit the network's ability to identify complex structures in images. Azad et al. have introduced the BDCU-Net architecture, combining a bidirectional ConvLSTM with U-Net for medical image segmentation \cite{azad2019bi}. This design optimizes model efficiency and complexity by utilizing dense convolutional layers in the final encoding path, thereby reducing the number of training parameters. In another paper, Zhang et al. proposed a novel model called BCU-Net for medical image segmentation \cite{zhang2023bcu}. BCU-Net fully analyzes both local details and overall context by combining the strengths of ConvNeXt and U-Net models. Also, this method tackles the issue of imbalanced detection of rare abnormalities by employing MRL loss, significantly improving diagnostic accuracy.  Yu et al. introduced EU-Net, Automatic U-Net Neural Architecture Search for Medical Image Segmentation Using Differential Evolutionary Algorithm \cite{yu2023eu}. EU-Net employs the DE algorithm to automatically optimize the U-Net neural network architecture. It leverages various strategies to effectively explore the search space and identify the optimal architecture for each specific dataset.  Zheng et al. introduced an end-to-end saliency detection network named HLU2-Net \cite{zheng2021hlu}. This network utilizes an innovative end-to-end structure, a hybrid loss function, and a coordinate attention module to combine and enhance features. In \cite{zulfiqar2023dru}, a neural network called DRU-Net is presented for the segmentation of pulmonary arteries, incorporating a uniquely crafted hybrid loss function. DRU-Net has shown enhanced capabilities in dealing with larger volumes of computed tomography images, effectively improving performance and mitigating overfitting. Furthermore, the efficacy of the devised hybrid loss function has been successfully implemented to boost the overall efficiency of the network in this study.

	Despite remarkable advancements in medical image segmentation, challenges still exist in deep learning for medical image segmentation. Main limitations are semantic gap problem, low learning efficiency, and insufficient information integration in skip connections. Attention mechanisms have emerged as a promising solution to address these challenges and enhance the performance of deep supervised learning methods for medical image segmentation \cite{jalali2024dabt}.

 \subsubsection{Attention guided models  } \label{sec:RelatedWork.deep.unet}
 
    Attention-based approaches enable models to selectively focus on specific regions or features of the input data that are most relevant for the task at hand. This can help to improve information integration in skip connections, reduce the problem of semantic gap, and improve learning efficiency. Using the attention mechanism in U-Net improves the ability of the network to focus on the critical areas of the image and increase the accuracy in the segmentation of the desired objects \cite{al2023improved}. Many researchers have used the attention mechanism in their segmentation network to improve model sensitivity and prediction accuracy in segmentation. Oktay et al. \cite{oktay2018attention}, proposed the Attention U-Net, which utilizes attention gates to highlight significant features in medical images. This approach contributes to more precise predictions in the segmentation of objects of interest. Yang et al. introduced NAUNet, a lightweight encoder-decoder network for retinal vessel segmentation in fundus images \cite{yang2023naunet}. It utilizes techniques such as DropBlock regularization, efficient attention modules, nested skip connections, data augmentation, and a mixed loss function to improve accuracy and address issues related to vascular images. Chen et al. proposed the Dual Attention Network (DANet) for multivariate time series classification \cite{chen2022net}. DANet utilizes two dual-attention layers to capture both local and global information in multivariate time series data. These layers enable the extraction of key features, thereby improving the classification performance. Chen et al. introduced an innovative method for medical image segmentation using the IBA-U-Net network \cite{chen2021iba}. This method incorporates three main ideas: integrating ConvLSTM and attention blocks to reduce semantic gaps, factoring convolutions using Redesigned Inception, and developing a deep convolutional neural network with multiscale features and incorporating an Attentive BConvLSTM mechanism. This approach not only improves segmentation accuracy but also achieves significant performance enhancements with lower computational costs compared to previous models. Wang et al. presented PCRTAM-Net, a deep learning model for automatically segmenting retinal vessels in color fundus images \cite{wang2023pcrtam}. The method incorporates three key elements: pre-activated dropout convolution residual for better feature learning, residual atrous convolution spatial pyramid for multiscale information extraction, and a triple attention mechanism for structural information capture. Evaluation on multiple datasets showed PCRTAM-Net achieving state-of-the-art results in retinal vessel segmentation, affirming its efficacy and advancement in this domain. Although the mentioned studies have achieved good results in medical image segmentation, they still face challenges such as low resistance to noise, low contrast in images, and complex structures.  
 
 \subsubsection{Feature fusion modules } \label{sec:RelatedWork.deep.fusion}
 
   In the medical field, the distinction and detection of sensitive areas and subtle changes in medical images are extremely importance. Feature fusion modules facilitate the extraction of crucial and fundamental information from different layers of a neural network, and by merging them, a unified and rich representation of image features is obtained \cite{zhang2020ifcnn}. This action facilitates the improved recognition of subtle differences in images and enhances detection accuracy. In scenarios where the identification and visualization of small and critical regions in medical images are imperative, the use of feature fusion modules in U-Net proves to be impactful and essential.

So far, some studies in this domain have focused on the development of models and architectures based on U-Net with feature fusion modules to enhance performance in lung segmentation. For example, Xie et al. \cite{xie2023multi} introduced a novel method for lung image segmentation. Their proposed approach utilizes the Multi-Scale Dilated Convolution (MSDC) feature fusion module to address challenges arising from low contrast and complex appearances in medical images. By incorporating this module into the network, the capability to analyze a broader range of information from images has improved, leading to a substantial increase in segmentation accuracy. The results demonstrate a significant improvement of this method compared to previous approaches in lung image segmentation. Jiang et al. \cite{jiang2023iu} have introduced a model called IU-Net for medical image segmentation. This model combines Swin Transformer and CNN, utilizing a feature fusion module based on wave function representation to enhance global contextual information and improve interaction with features. The feature fusion module, employing the wave function representation, integrates feature information from multiple sources, resulting in improved performance in skin lesion and lung field segmentation tasks.  Xu et al. \cite{xu2023lung} introduced a novel method for diagnosing breast and lung parenchyma radiological images using the Kiu-Net multi-interaction network model. This approach significantly improves feature fusion compared to existing methods and shows superior detection performance on the Montgomery County and Shenzhen datasets. The Feature Integration Module collects and merges features from U-Net and Ki-Net networks, enhancing accuracy and efficiency in radiological image detection by utilizing more comprehensive information. Sun et al. introduced a novel gland segmentation network called DARMF-UNet \cite{zhang2022automatic}, which improves gland segmentation accuracy compared to existing methods by utilizing deep supervision techniques and combining multi-scale features. DARMF-UNet incorporates modules such as Coordinate Parallel Attention (CPA) and Dense Atrous Convolution (DAC) to enhance semantic information in each layer and uses a hybrid loss function to improve segmentation accuracy. Experimental results demonstrate that this model outperforms existing methods in terms of segmentation accuracy and quality.

According to the proposed studies and inspired by the idea of Res-UNet, attention mechanism and feature fusion module, we considered a network based on u-net by integrating resnet blocks as encoder. We also designed a channel attention module and a feature fusion module.  Our network also uses a self refinement module. Finally, we decoded the modified features and then created a data modification network to improve. Also, our model uses a hybrid loss function. Therefore, our network is able to improve the accuracy and efficiency in detecting features in lung CT images. Also, the self refinement module allows the network to perform automatic feature refinement, and the combination of compression layers provides an overall improvement in model performance. In addition, the use of channel attention modules and feature fusion, together with the integration of resnet blocks, has helped to increase the accuracy and sensitivity in distinguishing important features and has increased the speed and efficiency of the network. The details of the proposed method are explained in detail in Section \ref{sec_Methodology}.

\section{Proposed Dataset} \label{sec:proposedDataset}
\subsection{Workflow Overview}
\begin{figure}[h]
	\centering
	\includegraphics[width=70mm]{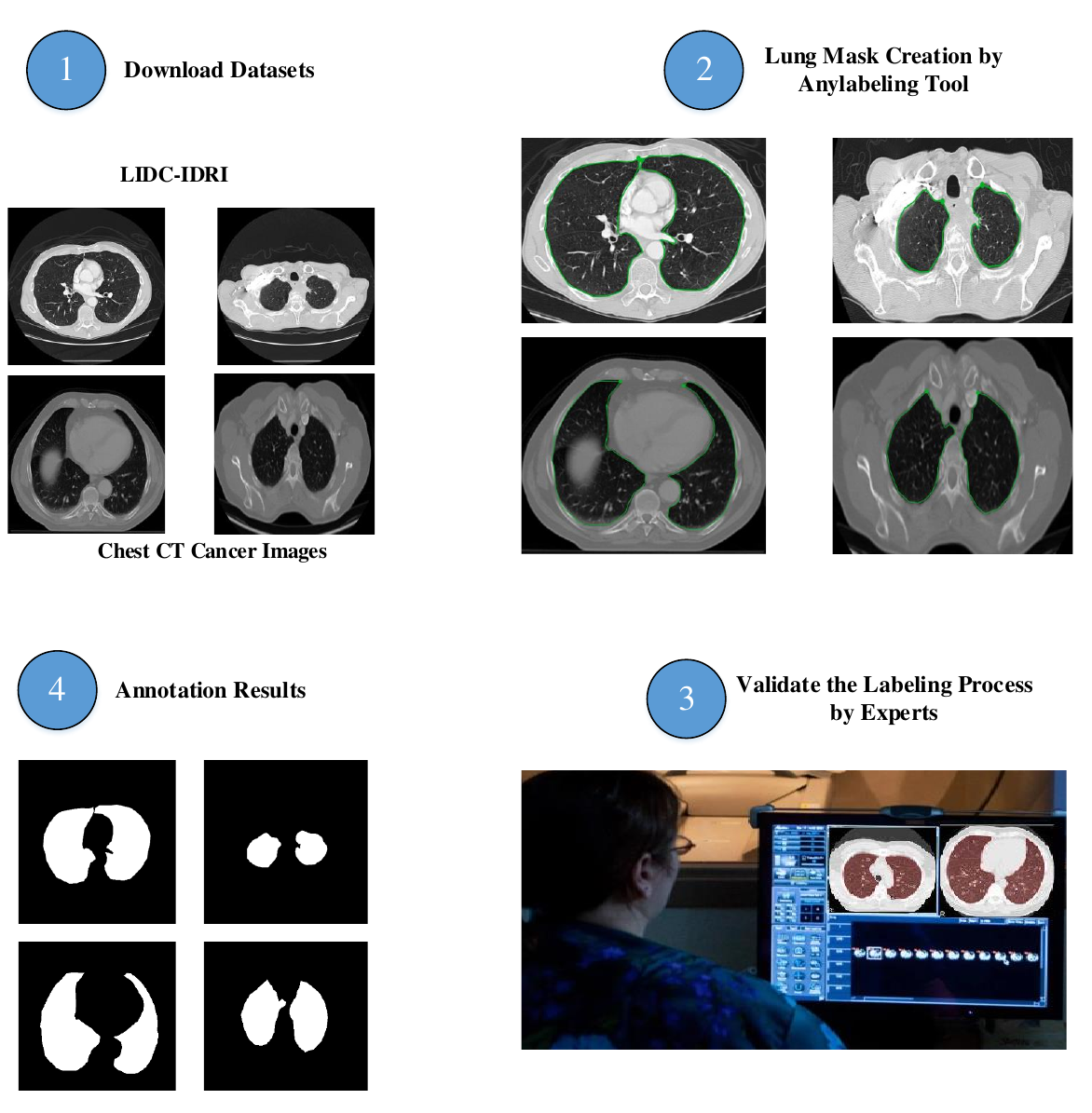}
	\begin{minipage}[]{3in}
		\caption{ Labeling lung CT images process.
			\label{fig_dataset_stages}
		}
	\end{minipage}
\end{figure}

In this paper, we introduce LungSegDB, a comprehensive dataset for lung segmentation. We followed a systematic workflow to prepare the datasets, which included downloading the datasets, creating lung masks using any labeling software, conducting an expert review for accurate validation, and finally evaluating the results of the annotation process. Figure \ref{fig_dataset_stages} illustrates the sequential steps of label generation. 
LungSegDB is our dataset, which is defined from two datasets: Version 1, which includes the LIDC-IDRI (Lung Image Database Consortium Image Collection) dataset \cite{armato2011lung}, from which we randomly selected 1800 DICOM images (referred to as Dataset Version 1), and Version 2, which includes the Chest CT-Scan Cancer Images Dataset from Kaggle  which we randomly chose 700 PNG/JPG images (referred to as Dataset Version 2).
 
Together, these datasets provided us with the raw CT images. Next, we created lung masks (segmentation labels) using the AnyLabeling software, which allowed us to accurately outline the lung regions in each CT image. After generating the masks, we conducted an expert review of the images to validate the labeling and ensure high-quality annotations for the lung segmentation task.

\subsection{Description of Datasets}
In this paper, we utilized two datasets for training and testing our lung segmentation model, collectively referred to as LungSegDB: Dataset Version 1, which is the LIDC-IDRI dataset, and Dataset Version 2, the Chest CT Cancer Images Dataset from Kaggle\footnote{\url{https://www.kaggle.com/datasets/dishantrathi20/ct-scan-images-for-lung-cancer}}. Below, we provide details on each dataset and the process of labeling the images.
   
Dataset Version 1 (LIDC-IDRI) consists of a collection of medical images used for the detection and analysis of lung nodules in CT scans. This dataset was created by the LIDC consortium and the National Institute of Biomedical Imaging and Bioengineering (IDRI). It includes 1018 cases from 1010 patients, with each case consisting of images from a clinical thoracic CT scan, accompanied by an associated XML file documenting the results from a two-phase image annotation process conducted by four experienced thoracic radiologists. The CT images are stored in DICOM (Digital Imaging and Communications in Medicine) format, featuring three channels and a resolution of 512 $\times$ 512. This dataset is publicly available, and researchers can access it through reputable medical imaging websites.

Since the images in the LIDC-IDRI dataset lack labels for lung tissue segmentation, which are crucial for training a segmentation model, we manually labeled these images and, for the first time, publicly released them along with their labels. Initially, we labeled 1800 images from Dataset Version 1.

To further enhance our training data, we incorporated an additional dataset, referred to as Dataset Version 2, the Chest CT Cancer Images Dataset from Kaggle. This dataset contains CT-scan images of patients with various types of chest cancer, including adenocarcinoma, large cell carcinoma, squamous cell carcinoma, as well as normal lung images. The images are stored in JPG and PNG formats and were initially collected for a chest cancer detection project. Since these images were also unlabeled for lung segmentation tasks, we manually labeled 700 images from this dataset.

In total, LungSegDB comprises 2,500 labeled images, combining 1800 images from Dataset Version 1 and 700 images from Dataset Version 2. For the training process, we selected 2,350 images, while the remaining 150 images were used for testing. This combined dataset enabled us to enhance the performance and generalization of our lung segmentation model.
\subsection{Lung Mask Creation with AnyLabeling Tool}
\begin{figure}[]
	\begin{center}
		\begin{tabular}{ccc}
			\includegraphics[width=2.2cm,height=2.2cm]{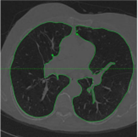}
			& \includegraphics[width=2.2cm,height=2.2cm]{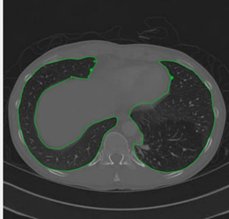}
			& \includegraphics[width=2.2cm,height=2.2cm]{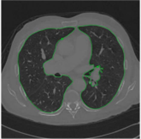}
		\end{tabular}
		\caption{Examples of the labeling extraction process from various images.
			\label{fig_pm_dataset}}
	\end{center}
\end{figure}

In this stage, we used the AnyLabeling  software to segment lung lobes and label the lung CT images. AnyLabeling is an advanced data labeling tool powered by artificial intelligence, designed for fast and efficient labeling. This software supports various types of labeling, including polygons, rectangles, circles, lines, points and etc. A standout feature of AnyLabeling is its use of advanced AI models such as YOLO and Segment Anything for automatic labeling. These models enable the software to perform initial labeling with high accuracy, which users can then manually refine and complete. Figure \ref{fig_pm_dataset} shows several examples of the labeling extraction process from different images. These images illustrate how the AnyLabeling.
\subsection{Expert Review and Generating Results}
After generating the initial labels using the AnyLabeling tool for all CT images, specialists in medical imaging and radiology thoroughly reviewed and approved them to ensure accuracy and reliability. These labels are now available through a \href{https://github.com/sadjadrz/Lung-segmentation-dataset}{Dataset link}, making them available for public use and future research. This approach not only speeds up the labeling process but also provides researchers and physicians with high-quality, precise data that can help improve disease diagnosis and treatment.
\subsection{challenges in Label generation}
Labeling lung regions in CT images presents numerous challenges that can significantly affect the accuracy of diagnoses. One of the most prominent issues is the presence of regions with varying density. High-density tissues, such as tumors, infections, or abscesses, may appear prominently white in CT images, complicating the decision-making process regarding their classification. It is often challenging to determine whether these high-density areas should be considered part of the lung or excluded from labeling. Conversely, lower-density regions, like air spaces within the lung, are typically classified as lung tissue. However, as diseases progress and increase tissue density, the boundaries become less distinct, making accurate labeling more difficult.
    
\begin{figure}[h]
	\centering
	\includegraphics[width=80mm]{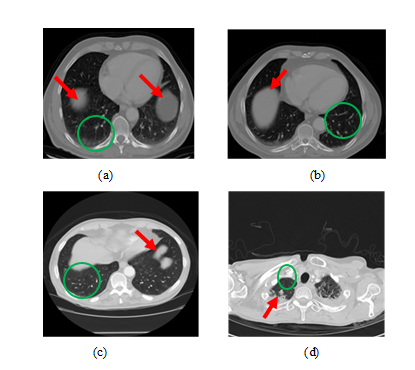}
	\begin{minipage}[]{3in}
		\caption{ Examples of the labeling extraction process from various images
			Images (a) and (b) correspond to Dataset Version 1 (LIDC-IDRI), while images (c) and (d) correspond to Dataset Version 2 (Chest CT Cancer Images Dataset).
			\label{fig_dataset_challanges}
		}
	\end{minipage}
\end{figure}

The primary criterion for labeling the lungs in CT images is the accurate identification of the anatomical boundaries of the lungs. These boundaries are typically distinguishable from other structures (such as the chest wall, diaphragm, and heart). The lungs appear as relatively organized, air-filled structures in CT images, showing up in dark color (low density). Additionally, lung tissue includes blood vessels and bronchi, which appear in CT images as thin and brighter structures (white or light gray dots). These areas represent the airways and blood vessels within the lungs and are considered part of the normal lung tissue. Examples of these areas are depicted in Figure \ref{fig_dataset_challanges} with green circles.
Areas with higher density, which usually appear bright (white) in CT images, may include tumors, infections, or other abnormal structures that do not appear darker than normal lung tissue due to their high density. These areas are typically identified as abnormal or pathological, with examples highlighted in Figure 3 using red arrows in each image.

The challenge arises in labeling lung CT images, as even if high-density areas (such as cancerous masses or inflammation) exist within the lung, these areas are still considered part of the lung structure. If a cancerous mass or any lesion with higher density is present in the lung (appearing as large white spots in CT images), this area may be recognized by the system as non-lung and ignored during labeling. This highlights the importance of accuracy in labeling and the need for advanced methods to correctly identify lung regions.

In this regard, we have been able to identify these challenges and accurately label the lung areas with precision and attention to detail. Our work is valuable as this accuracy in labeling can help improve the diagnosis and treatment of lung diseases, positively impacting the overall quality of healthcare services.
By addressing these challenges in our labeling process, we emphasize the significance and value of our work in accurately labeling lung regions in CT images, ultimately contributing to better disease diagnosis and treatment.
\section{Methodology}\label{sec_Methodology}
\begin{figure*}[h]
	\centering
	\includegraphics[width=185mm]{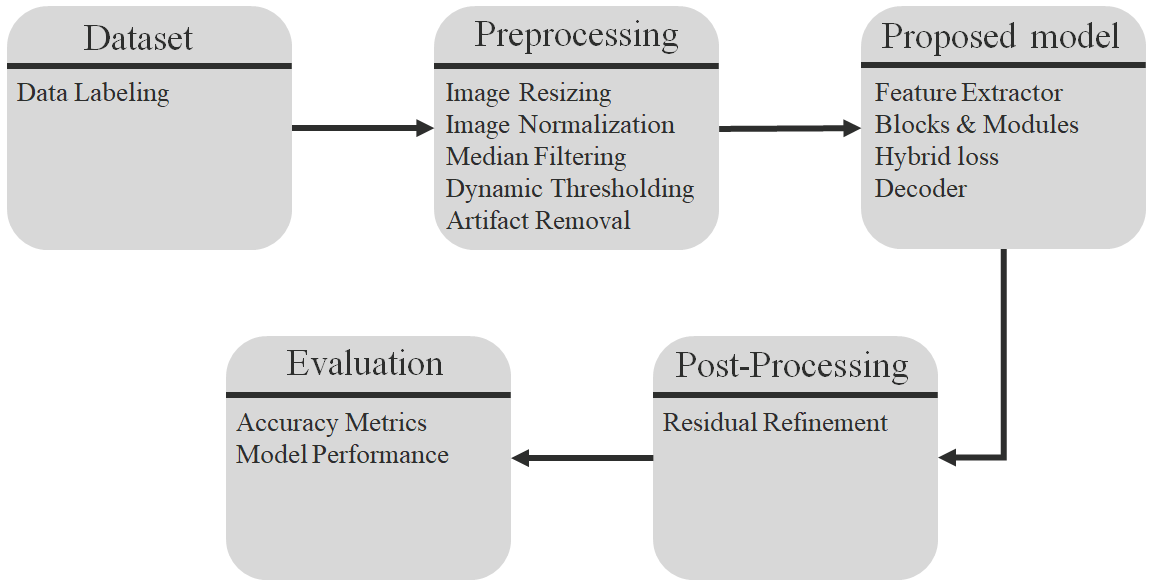}
	\begin{minipage}[]{6in}
		\caption{ Overall structure of pipeline of proposed framework.
			\label{fig_pm_pip}
		}
	\end{minipage}
\end{figure*}

Our suggested approach comprises five integrated phases within the general pipeline for lung segmentation, illustrated in Figure \ref{fig_pm_pip}. The process begins with the initial preparation of the dataset, where data labeling is essential for accurate image analysis. Following this, the images enter the preprocessing stage, which includes several critical steps to standardize and enhance the input data. Initially, each image is standardized through resizing to ensure uniformity across the dataset. Image normalization is applied to scale pixel values, followed by noise reduction through median filtering, which preserves important edges. Dynamic thresholding enhances the contrast to highlight important features while suppressing irrelevant background. Additionally, artifact removal is performed to eliminate unwanted components from the images. Following preprocessing, the images are fed into the proposed network, FusionLungNet. The final stage involves a thorough evaluation of the segmented images using a set of accuracy metrics and model performance assessments.

This section begins by providing a comprehensive overview of FusionLungNet. Then, we present in detail the CAA module, which is used to enhance the feature representation in the network. Next, we introduce the MFF block strategically operating on different layers of the decoder to enhance the information correlation between samples. Subsequently, our focus turns to the design of the Residual Refinement Module, dedicated to refining image boundaries. In addition, we use self refinement block, enhancing features through self-modality considerations, facilitating an efficient transfer into the decoder. Finally, we explain the network training loss. The integration of these modules enables FusionLungNet to achieve excellent performance in lung segmentation tasks.

In the final part of this section, we detail the preprocessing stage, essential for preparing the dataset for effective segmentation. This stage encompasses critical techniques such as image resizing, median filtering, dynamic thresholding and artifact removal.
\subsection{Overview of the Proposed Network} \label{sec_Methodology_Overwiew}
\begin{figure*}[h]
	\centering
	\includegraphics[width=185mm]{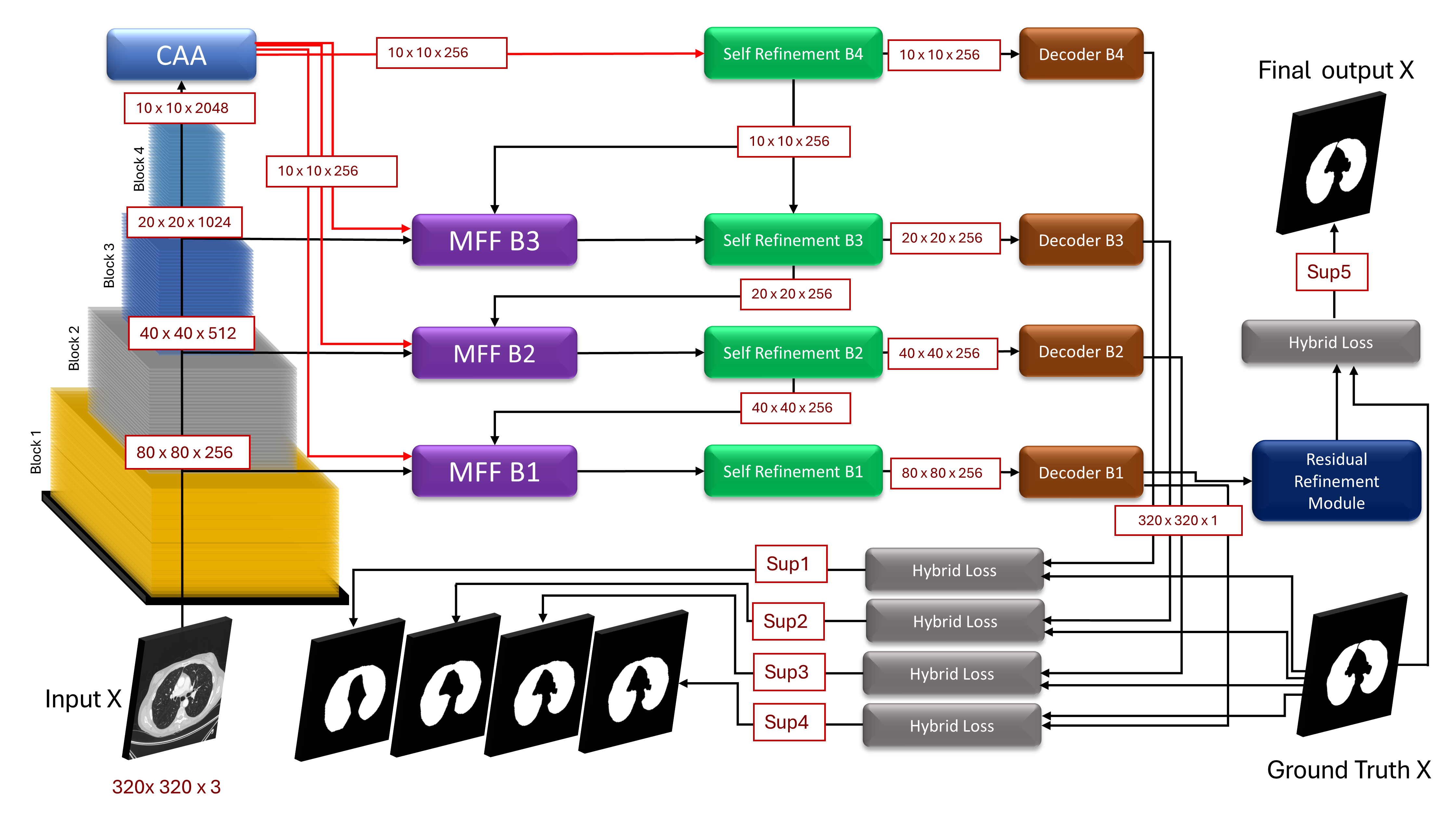}
	\begin{minipage}[]{6in}
		\caption{ Architecture of our proposed FusionLungNet.
			\label{fig_pm_x}
		}
	\end{minipage}
\end{figure*}

FusionLungNet is a multi-level hybrid end-to-end model, taking images as input and producing images as output to predict binary maps for segmentation task. The overall architecture of our FusionLungNet is shown in Figure \ref{fig_pm_x}. In the first step, images  are fed into the ResNet-50 backbone network to capture contextual information essential for extracting target features. The obtained features through this process are organized into four layers, $\{ \text{Block}\quad i \,|\, i = 1, 2, 3, 4 \}$. Then the highest-level features (output of  \text{Block} 4) are input into CAA module to improve the accuracy of the extracted target features. 

Low-level features capture fine details such as texture and boundaries but may include extra background noise. On the other hand, high-level features provide abstract information, helping identify important objects and reduce noise. Additionally, extracting more global features can be advantageous for various tasks, as these features offer a broader context and contribute to better overall performance. As a result, it's common to combine these two levels of features to achieve a more comprehensive representation. However, this combination suffers from some defects due to the contradictory response of different layers. To address this, we employ a self refinement module and integrate it with the two-level features generated by CAA and different encoding parts of the model. This integration helps alleviate the challenges associated with multi-scale feature fusion modules.

Finally, through the up-sampling process, output maps with the same resolution as the original image are obtained. Throughout the training process at various stages, we employ a hybrid loss to supervise the training for accurate predictions. Additionally, to enhance the final binary map produced by the prediction module, we incorporate a Residual Refinement Module that employs a residual encoder-decoder architecture.
\begin{figure*}[t]
	\centering
	\includegraphics[width=185mm]{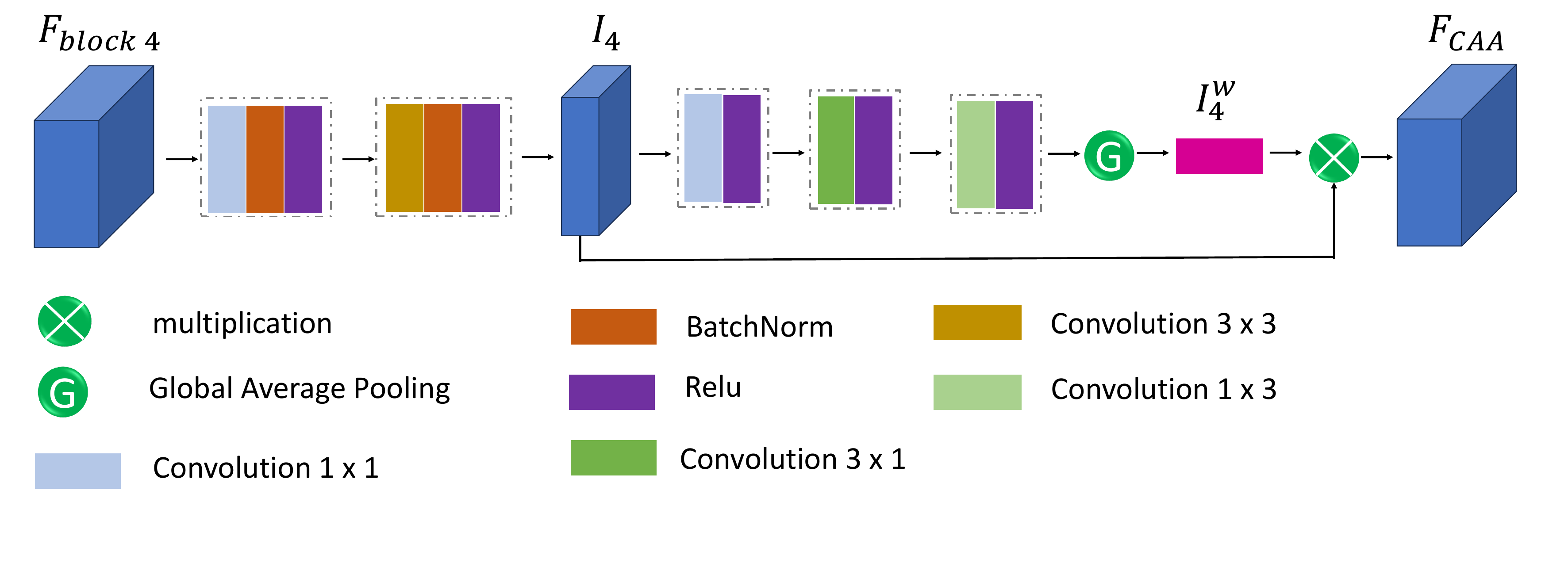}
	\begin{minipage}[]{6in}
		\caption{ Schematic of the Channel-Wise Aggregation Attention  module.
			\label{fig_pm_CAA}
		}
	\end{minipage}
\end{figure*}
\subsection{Channel-Wise Aggregation Attention Module (CAA)} \label{sec_Methodology_CAA}
In this study, we introduce the Channel-Wise Aggregation Attention (CAA) module as a crucial element for enhancing the discriminative power of high-level features, specifically $F_{\text{block4}}$, within our deep learning architecture for lung segmentation. The CAA module is meticulously designed to refine feature representations by emphasizing important channels while suppressing less relevant ones, thereby improving the network’s focus on critical spatial information.
\subsubsection{Architecture and Operations}
As depicted in Figure \ref{fig_pm_CAA}, the CAA module integrates several key operations aimed at optimizing feature maps with minimal computational overhead:
\begin{itemize}
	\item \textbf{Initial Convolutional Processing}: 
	The module begins with a sequence of convolutional layers—first a \( 1 \times 1 \) convolution followed by a \( 3 \times 3 \) convolution. The \( 1 \times 1 \) convolution serves to reduce the dimensionality of the feature maps, facilitating efficient computation, while the subsequent \( 3 \times 3 \) convolution captures spatial dependencies and extracts common information across channels. This dual convolutional approach generates an intermediate feature map \( I_4 \), as defined in Equation~\eqref{eq:I4}:
	
	\begin{equation}
		I_4 = C_{3 \times 3}\left(C_{1 \times 1}\left(F_{\text{block4}}\right)\right)
		\label{eq:I4}
	\end{equation}
	
	Where, \( F_{\text{block4}} \) represents the input feature map, and \( C \) denotes the convolution operation.
	
	\item \textbf{Asymmetric Convolutions for Enhanced Contextual Encoding}:
	To further enrich the feature representation, the CAA module employs asymmetric convolutions—specifically \( 1 \times 3 \) and \( 3 \times 1 \) convolutions. Inspired by the findings in \cite{szegedy2016rethinking}, asymmetric convolutions are advantageous for capturing contextual information from varying receptive fields, especially effective for small-sized feature maps typical in high-level layers. These convolutions enable the network to assimilate multi-scale contextual cues without significantly increasing computational complexity.
	
	\item \textbf{Global Average Pooling (GAP) for Channel Weight Calculation}:
	Following the asymmetric convolutions, GAP is applied to aggregate spatial information, producing channel-wise statistics. This operation calculates the average activation for each channel, yielding the channel weights \(  I_{4}^ {w} \) as per Equation~\eqref{eq:I4w}:
	
	\begin{equation}
		 I_{4}^ {w} = \text{GAP}\left(C_{3 \times 1}\left(C_{1 \times 3}\left(I_4\right)\right)\right)
		\label{eq:I4w}
	\end{equation}
	
	These weights effectively serve as attention scores, quantifying the importance of each channel in the context of the entire feature map.
	
	\item \textbf{Channel-Wise Attention Mechanism}:
	The final step involves modulating the original feature map \( I_4 \) with the computed attention weights \(  I_{4}^ {w} \). This is achieved through element-wise multiplication, resulting in the refined feature map \( F_{\text{CAA}} \) as described in Equation~\eqref{eq:FCAA}:
	
	\begin{equation}
		F_{\text{CAA}} = I_4 \odot  I_{4}^ {w}
		\label{eq:FCAA}
	\end{equation}
	
	Where, \( \odot \) denotes element-wise multiplication. This operation ensures that channels with higher attention scores contribute more significantly to the subsequent layers, thereby enhancing the network’s ability to focus on salient features relevant for accurate lung segmentation.
\end{itemize}
\subsubsection{Rationale and Benefits}

The incorporation of the CAA module addresses several critical aspects in deep learning-based segmentation:

\begin{itemize}
	\item \textbf{Enhanced Feature Discrimination}: By assigning adaptive weights to each channel, the CAA module allows the network to prioritize features that are more pertinent to lung structures, thereby improving the overall segmentation accuracy.
	
	\item \textbf{Efficient Computational Utilization}: The strategic use of \( 1 \times 1 \) and asymmetric convolutions ensures that the module remains computationally efficient, making it suitable for real-time or resource-constrained applications.
	
	\item \textbf{Contextual Awareness}: The combination of asymmetric convolutions and GAP facilitates a comprehensive understanding of the feature maps’ spatial and channel-wise contexts, enabling more nuanced feature refinement.
\end{itemize}

In summary, the CAA module plays a crucial role in our network by effectively harnessing channel-wise attention to optimize high-level feature representations. This leads to more precise and computationally efficient lung segmentation, addressing both the accuracy and efficiency requirements of medical imaging applications.
\subsection{Multi-scale Feature Fusion block (MFF)} \label{sec_Methodology_MFF}
Effective image segmentation requires the integration of features from multiple scales to capture both the high-level semantic context and the low-level spatial details. High-level features provide extensive semantic information but often lack precise spatial resolution, while low-level features offer detailed spatial information but may include noise and lack contextual understanding. To harness the strengths of both, we propose the MFF block, designed to seamlessly integrate features across different scales and levels of abstraction.

As illustrated in Figure \ref{fig_pm_MFF}, the MFF block fuses three input feature maps: $I_1$, $I_2$, and $I_3$. Specifically: \begin{itemize} \item $I_1$ is the output from the CAA module, capturing global contextual information through channel attention mechanisms. \item $I_2$ is the feature map from the encoder block, containing high-level semantic features with reduced spatial resolution. \item $I_3$ is derived from the self refinement block (detailed in subsection \ref{sec_Methodology_self}), which enhances feature representations by refining and reinforcing important details. \end{itemize}

To prepare for fusion, each input feature map undergoes a $3 \times 3$ convolution operation, denoted as $C_{3\times3}$, to extract salient features and unify the number of channels across all inputs. To ensure spatial alignment, an up-sampling operation, represented by $R$, is applied to $I_1$ and $I_3$ so that all feature maps match the spatial resolution of $I_2$ from the encoder.

The core fusion process involves pairwise element-wise multiplication of the adjusted feature maps, capturing the complementary information between different scales:
\begingroup\makeatletter\def\f@size{10}\check@mathfonts
\begin{align}\label{eqmff}
	& F_{ab} = C_{3\times3}(R(I1)) \odot C_{3\times3}(I2)\\
	& F_{ac} = C_{3\times3}(R(I1)) \odot C_{3\times3}(R(I3))\\
	& F_{bc} = C_{3\times3}(I2) \odot C_{3\times3}(R(I3))\\
	& F_{MFF} = C_{3\times3}(concatenation(F_{ab},F_{ac},F_{bc}))
\end{align}
Where,  \( \odot \) denotes element-wise multiplication, which effectively emphasizes the shared important features between the paired inputs while suppressing less relevant information.

The resulting feature maps $F_{ab}$, $F_{ac}$, and $F_{bc}$ are then concatenated to aggregate the multi-scale information: \begin{align} & F_{MFF} = C_{3\times3}(\text{concatenate}(F_{ab}, F_{ac}, F_{bc})) \end{align} A final $3 \times 3$ convolutional layer further refines the fused features, enhancing the model's ability to focus on critical aspects of the input data by combining rich semantic context with detailed spatial information.

By integrating features from multiple scales through these operations, the MFF block effectively captures and consolidates information across different levels of the network. This multi-scale fusion enhances the segmentation performance by providing a more comprehensive understanding of the image content, allowing the model to make more accurate predictions.

\begin{figure*}[h]
	\centering
	\includegraphics[width=185mm]{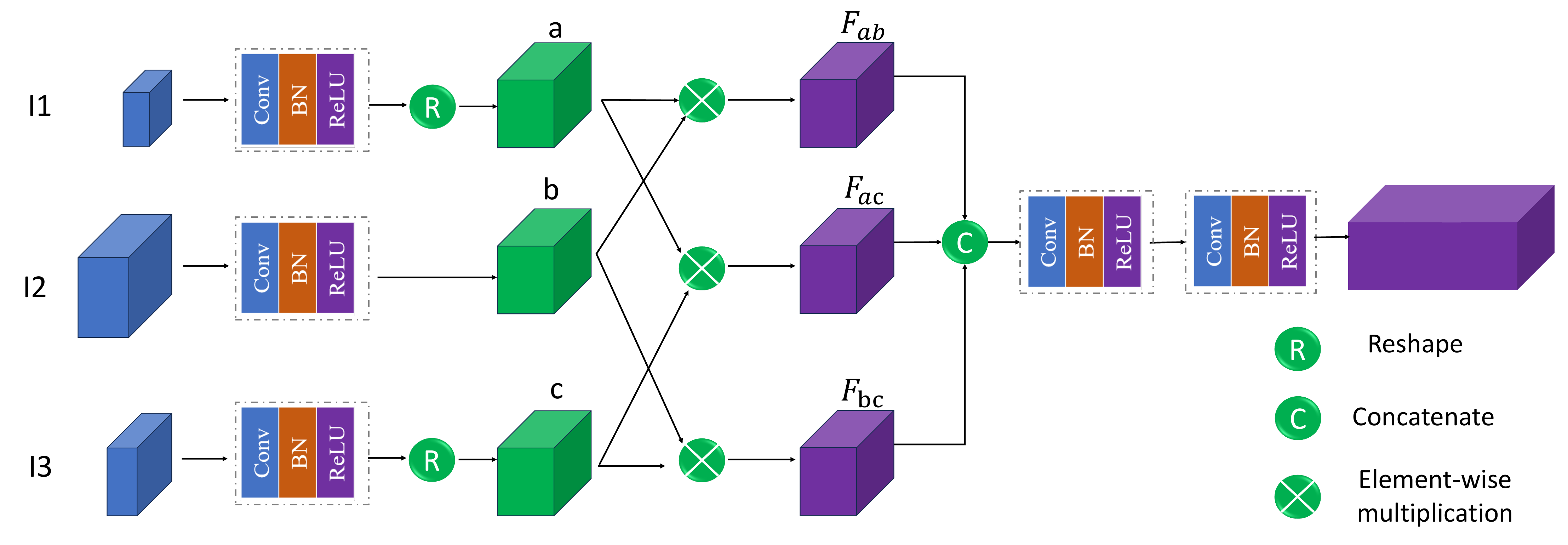}
	\begin{minipage}[]{6in}
		\caption{ Detailed network architecture diagram of The multi-scale feature fusion block.
			\label{fig_pm_MFF}
		}
	\end{minipage}
\end{figure*}
\subsection{Self Refinement Block} \label{sec_Methodology_self}
Drawing inspiration from \cite{chen2020global}, the integration of the Self Refinement (SR) Block successive to each MFF block is crucial for improving and enhancing feature mapping, mitigating the defects associated with the direct fusion of different layers, and ensuring the generation of more discriminative features for more accurate saliency maps. As shown in Figure \ref{fig_pm_FFM},  Initially, we employ a 3 $\times$ 3 convolutional layer to compress the input features, denoted as $F_{\text{in}}$, into a feature vector $F_{\text{1}}$ with a channel dimension of 256, while retaining valuable information. In other direction, 256 channels expanded to 512 by convolution 3 $\times$ 3 ($F_{\text{2}}$). Following this, the 512 channels are split into two segments, each containing 256 channels. This division results in the generation of $F_{\text{m}}$ for the multiplication operation and $F_{\text{c}}$ for the addition operation. 

\begin{figure*}[h]
	\centering
	\includegraphics[width=185mm]{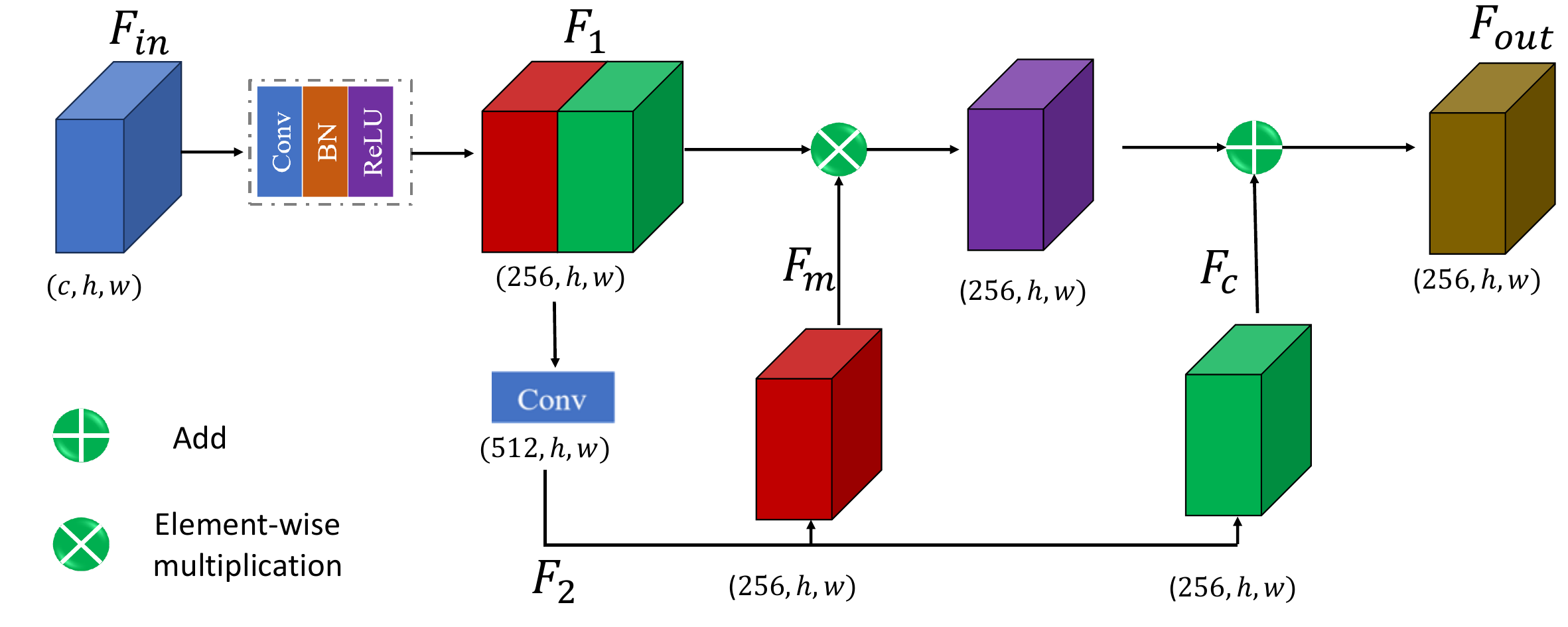}
	\begin{minipage}[]{6in}
		\caption{ Details of the proposed self refinement block.
			\label{fig_pm_FFM}
		}
	\end{minipage}
\end{figure*}
\begin{figure*}[h]
	\centering
	\includegraphics[width=185mm]{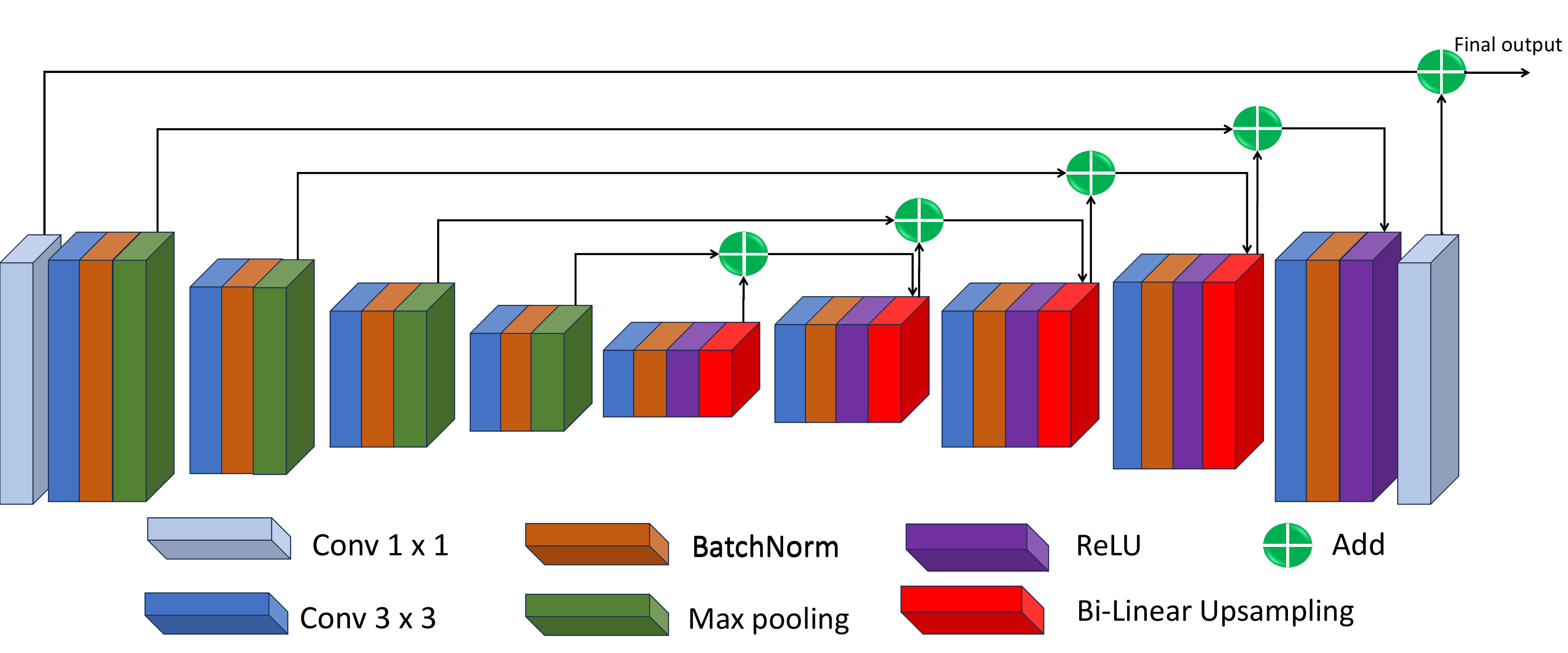}
	\begin{minipage}[]{6in}
		\caption{ Architecture of our proposed Residual refinement module.
			\label{fig_pm_RRM}
		}
	\end{minipage}
\end{figure*}
\subsection{Residual Refinement module} \label{sec_Methodology_segmentation network}
The output from the decoder part fails to accurately segment the boundary between the lung and non-lung areas. To address this limitation and enhance boundary delineation, we introduce a Residual Refinement module. The Residual Refinement Module adopts a residual encoder-decoder architecture to mitigate the boundary identification challenges in the segmentation process.

As shown in Figure \ref{fig_pm_RRM}, our approach initiates with a 3 $\times$ 3 convolution. Subsequently, a sequence of five convolutional layers and four pooling layers is employed to encode the feature map. In the decoder part, resolution recovery is achieved through bi-linear interpolation, and essential high-resolution features are complemented by leveraging shortcut connections that bridge the contracting and expansive paths. At the final layer, we employ a 1  $\times$ 1 convolution to generate the ultimate binary segmentation result.

\subsection{Hybrid Loss} \label{sec_Methodology_loss function}
We propose a hybrid loss function, comprising focal loss, SSIM loss, and IoU loss, to enhance regional segmentation and boundary accuracy(Eq.\ref{lbl_loss}).
\begingroup\makeatletter\def\f@size{10}\check@mathfonts
\begin{flalign} \label{lbl_loss}
	& {Loss} = \lambda_{1} Loss_{focal} + \lambda_{2} Loss_{ssim} + \lambda_{3} Loss_{iou}
\end{flalign}
\endgroup
where $\lambda$ denotes the weight of different loss, and According to the quantitative comparison in Table \ref{fig_alpha}, we set \({\lambda}_{1} = 0.3 \), \({\lambda}_{2} = 0.4 \), \({\lambda}_{3} = 0.3 \).

The total loss comprises two components: the primary loss linked to the Residual Refinement Module output and the supplementary loss from each decoder.
\begingroup\makeatletter\def\f@size{10}\check@mathfonts
\begin{flalign}
	& {Loss}_{total} = Loss_{prim} + \sum_{n=1}^{4} { Loss_{sup}^ {n} }  
\end{flalign}
\endgroup
$Loss_{prim}$ and $Loss_{sup}$ indicate the primary and supplementary losses, respectively. The supplementary losses are only present during the training stage and are discarded during inference.

In proposed dataset, the segmented lung regions within certain images constitute a relatively small proportion of the entire image, as illustrated in Figure \ref{fig_pm_focall}. This class imbalance highlights the need for specialized techniques, such as the application of focal loss, to address the inherent challenges associated with binary lung segmentation. Specifically, focal loss is employed to mitigate the class imbalance resulting from uneven distributions of foreground and background pixels. By assigning greater importance to the lung region during training, focal loss enhances the model's capability to accurately converge on all pixels. It is defined as:
\begin{align}
	& L_{binary-focal}=
	\left\{\begin{matrix}
		-\alpha(1-y')^{\gamma}\log(y'), \ \ \ y=1\\
		-(1-\alpha)(y')^{\gamma}\log(1-y'), \ \ \ y=0\\
	\end{matrix}\right.
\end{align}

\begin{figure*}[]
	\begin{center}
		\begin{tabular}{ccccc}
			\includegraphics[width=2.2cm,height=2.2cm]{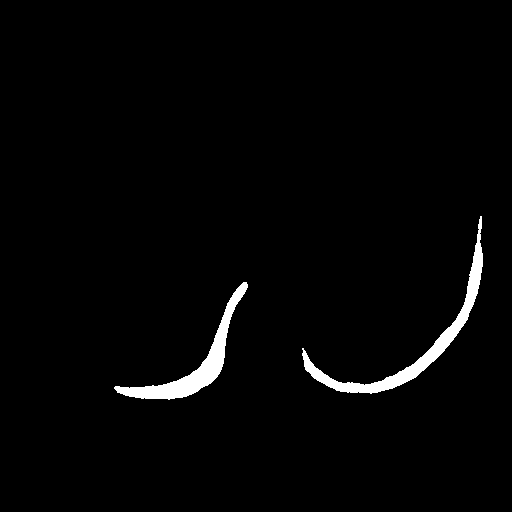}
			& \includegraphics[width=2.2cm,height=2.2cm]{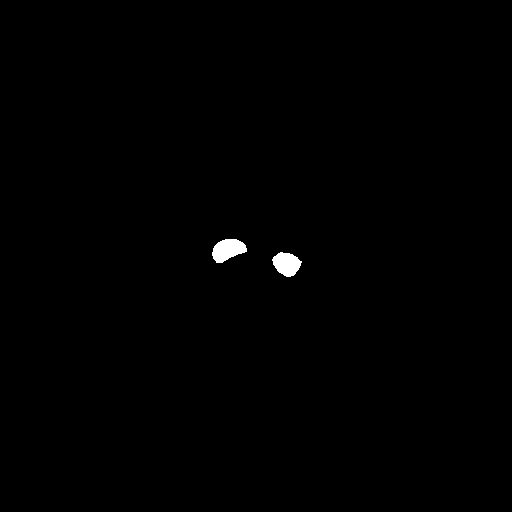}
			& \includegraphics[width=2.2cm,height=2.2cm]{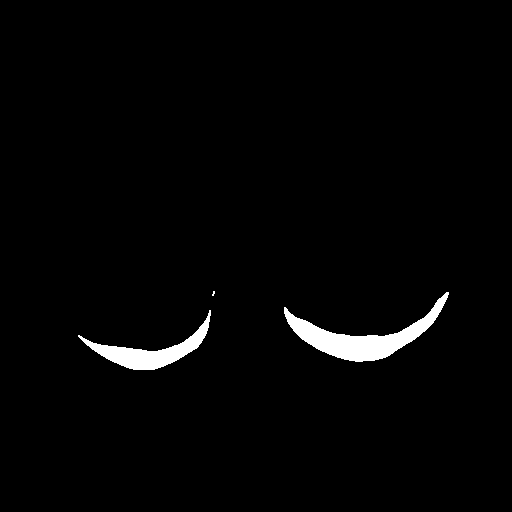}
			& \includegraphics[width=2.2cm,height=2.2cm]{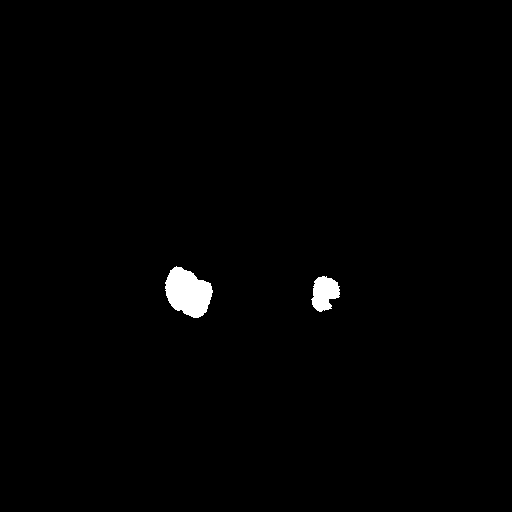} 
			& \includegraphics[width=2.2cm,height=2.2cm]{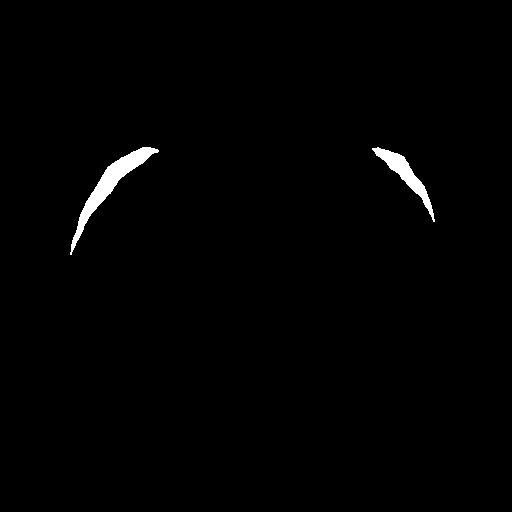} 
		\end{tabular}
		\caption{ Demonstrates the challenge of class imbalance in our dataset, with segmented lung regions forming a small proportion of the image.
			\label{fig_pm_focall}}
	\end{center}
\end{figure*}
\begin{table*}[h]
	\centering
	\begin{tabular}{|c|c|c|c|c|c|}
		\hline
		$\lambda_1 \rightarrow loss_{focal}$ & $\lambda_2 \rightarrow loss_{SSIM}$ & $\lambda_3 \rightarrow loss_{IoU}$ & IoU   & $F1$-Score & Accuracy \\ \hline
		0.3 & 0.4 & 0.3 & \textbf{98.04} & \textbf{99.02} & \textbf{99.73} \\ \hline
		0.4 & 0.3 & 0.3 & 97.86 & 98.6 & 99.65 \\ \hline
		0.3 & 0.3 & 0.4 & 97.73 & 98.53 & 99.63 \\ \hline
	\end{tabular}
		\caption{ Quantitative comparison of $\lambda$ for our hybrid loss.
	\label{fig_alpha}
}
\end{table*}
Utilizing the ssim loss \cite{cai2021multi} as defined in Eq. \ref{SSMI}, we focus on the patch level to capture structural data and enhance detailed boundary predictions. This approach involves considering adjacent neighbourhoods for individual pixels, assigning increased weight to edges, thereby optimizing the segmentation of boundaries.
\begin{align}
	ssim = 1 - \frac{(2\mu_x\mu_y + C_1) + (2 \sigma _{xy} + C_2)} 
	{(\mu_x^2 + \mu_y^2+C_1) (\sigma_x^2 + \sigma_y^2+C_2)}
	\label{SSMI}
\end{align}
The covariance of $x$ and $y$ is represented by $\sigma_{xy}$. The mean and standard deviation of $x$ and $y$ are represented by $\mu_x$, $\mu_y$, $\sigma_x$, and $\sigma_y$, respectively. In order to prevent division by zero, we empirically fixed $\mathbf{C}_1 = 0.012$ and $\mathbf{C}_2 = 0.032$ from this paper \cite{wang2003multiscale}.

\subsection{Preprocessing}\label{sec_preprocessing}
Preprocessing is the first and essential step in automatic lung segmentation. A common limitation in lung imaging is the frequent suboptimal quality of the CT scans due to noise, low contrast, and irrelevant background structures. To achieve segmentation with high accuracy, our objectives are twofold: firstly, to enhance the quality of the CT images, and secondly, to focus on the lung regions by eliminating unnecessary background information. Our preprocessing pipeline aims to improve the clarity and detail of the lung images through several techniques. The preprocessing consists of image resizing, median filtering, dynamic thresholding and lung region segmentation.
The initial step involves resizing all CT images to a standardized dimension of 320 $\times$ 320 pixels. This standardization is crucial for maintaining uniformity across the dataset. 

After resizing, we apply median filtering to each image to reduce noise while preserving important anatomical details. CT images often contain artifacts that can obscure fine structures within the lungs. Median filtering effectively mitigates this issue by replacing each pixel's value with the median value of its neighboring pixels. This process smooths the image without blurring critical edges and features essential for accurate segmentation.

To enhance the contrast of the images and emphasize the lung tissues, we employ contrast enhancement techniques based on dynamic thresholding. Contrast enhancement is important in improving the visibility of structures within medical images, which can significantly impact the performance of segmentation algorithms. Recent studies have demonstrated the effectiveness of adaptive contrast enhancement methods in medical imaging applications \cite{vijayalakshmi2022novel,vijayalakshmi2023systematic,vijayakumar2025sustainable,rezvani2024single}. We adopt a dynamic thresholding approach inspired by these recent advancements. Unlike global thresholding methods that apply a single threshold value to the entire image, dynamic thresholding calculates threshold values based on local image characteristics. This adaptive approach allows for better differentiation between lung tissues and surrounding structures, especially in images with uneven illumination or varying contrast levels.

The dynamic threshold $T(x, y)$ at each pixel location $(x, y)$ is computed using local statistics within a neighborhood window. The enhanced image $I_{\text{enh}}(x, y)$ is then obtained by applying a transformation function that adjusts the pixel intensities based on the local threshold:

\begin{align}
	I_{\text{enh}}(x, y) = 
	\begin{cases}
		\alpha \cdot I_{\text{med}}(x, y) + \beta, & \text{if } I_{\text{med}}(x, y) > T(x, y) \\
		I_{\text{med}}(x, y), & \text{otherwise}
	\end{cases}
\end{align}

where $I_{\text{med}}(x, y)$ is the median-filtered image, and $\alpha$ and $\beta$ are parameters controlling the level of enhancement. This method enhances the contrast of lung tissues while suppressing irrelevant background, improving the subsequent segmentation process.

The final step in the preprocessing pipeline is artifact removal. After enhancing the images and suppressing background noise, we eliminate unwanted artifacts that may interfere with accurate lung segmentation. Artifacts, such as scanner-induced distortions or irrelevant structures, are removed to ensure that only meaningful information remains in the image. This process helps reduce interference from non-lung regions, allowing the segmentation model to focus solely on relevant anatomical structures. By effectively removing artifacts, we improve the quality of the input data, enhancing the efficiency and accuracy of the segmentation model.

To demonstrate the effectiveness of our preprocessing steps, we visually present the transformation of a sample CT image through each stage in Figure \ref{fig_pm_preprocessing}. This figure illustrates how each preprocessing technique contributes to enhancing the image quality and isolating the lung region for accurate segmentation.
\begin{figure*}[h]
	\centering
	\includegraphics[width=130mm]{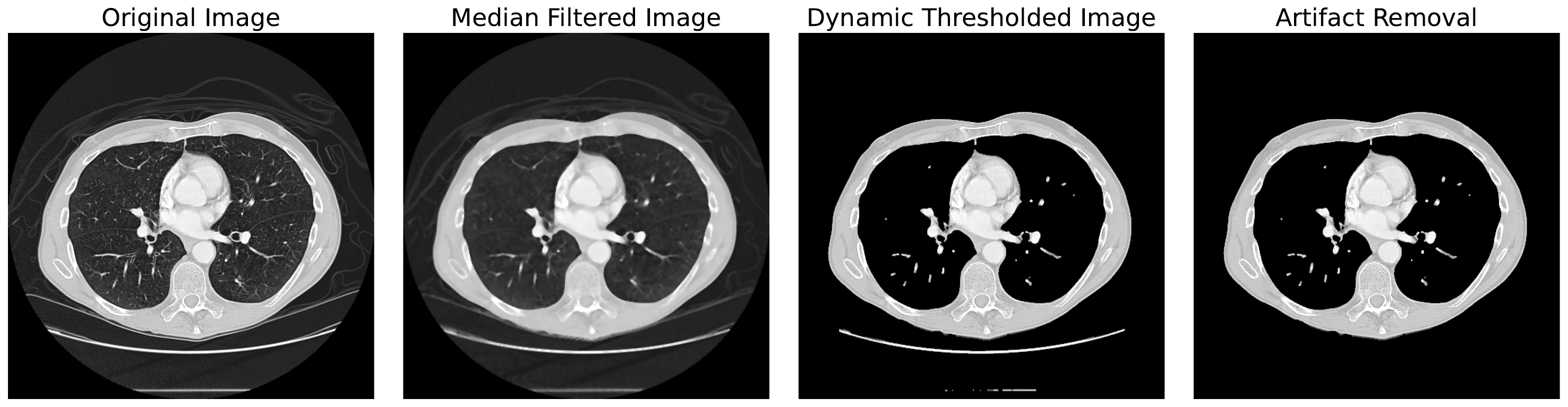}
	\begin{minipage}[]{6in}
		\caption{ Preprocessing steps in the proposed lung segmentation method.
			\label{fig_pm_preprocessing}
		}
	\end{minipage}
\end{figure*}
\section{Experiments}\label{sec_Experiments}
In this section, we performed experiments to compare the performance of the proposed lung segmentation dataset with state-of-the-art deep learning models for segmentation. In the training process, we implemented our model in two scenarios.

Secondly, we trained the model on both LungSegDB-V1 and LungSegDB-V2 after applying the preprocessing techniques discussed in Section \ref{sec_preprocessing}. We expanded the dataset to include 2,500 labeled images, which were divided into a training set (2,350 images) and a test set (150 images).

\begin{table*}[h]
	\begin{center}
			\resizebox{\columnwidth*2}{!}{
			\begin{tabular}{lcccccc}
			 \toprule
			\textbf{Model} & \textbf{IoU} & \textbf{F1-score} & \textbf{Precision} & \textbf{Recall} & \textbf{Acc} & \textbf{MCC} \\
			\midrule
			Unet \cite{ronneberger2015u}          & 95.02  & 96.22  & 96.11 & 96.34 & 95.18 & 94.68 \\
			RU-Net \cite{alom2018nuclei}         & 94.93  & 96.35  & 95.52 & 97.21 & 97.15 & 94.80 \\
			ResNet34-Unet \cite{lau2020automated}  & 95.28  & 97.83  & 97.32 & 98.35 & 96.73 & 96.87 \\
			BCDU-Net \cite{azad2019bi}      & 96.32  & 98.52  & 99.02 & 98.03 & 97.21 & 97.61 \\
			ResBCDUnet \cite{jalali2021resbcdu}     & 97.15  & 98.05  & \textbf{99.12} & 97.01 & 97.58 & 96.60 \\
			NasNet \cite{zhang2022automatic}        & 97.56  & 98.71  & 98.93 & 98.45 & 98.76 & 97.90 \\
			DABT-U-Net \cite{jalali2024dabt}     & 97.31  & 98.79  & 98.24 & 98.86 & 98.45 & 97.90 \\
			ABANet  \cite{rezvaniabanet}       & 97.57  & 98.55  & 98.70 & 98.30 & 99.34 & 97.63 \\
			\midrule
			\textbf{Ours}  & \textbf{98.04} & \textbf{99.02} & 98.65 & \textbf{99.41} & \textbf{99.73} & \textbf{98.67} \\
			\bottomrule
					\caption{ A quantitative comparison of the results obtained on the first version of proposed lung dataset test set using a variety of techniques for image segmentation. The best scores are highlighted with bold.
						\label{tbl_exp_imgsegmentation} 
					}
			\end{tabular}
		}
	\end{center}
\end{table*}%
\begin{table*}[h]
	\begin{center}
		\resizebox{\columnwidth*2}{!}{
			\begin{tabular}{lcccccc}
				\toprule
				Model & IoU & F1-score & Precision & Recall & Acc & MCC \\
				\midrule
				Unet \cite{ronneberger2015u}           & 95.62  & 96.93  & 97.28 & 96.74 & 98.24 & 95.93 \\
				BCDU-Net  \cite{azad2019bi}     & 97.56  & 98.85  & 98.49 & 98.95 & 98.67 & 98 \\
				ABANet  \cite{rezvaniabanet}       & 97.57  & 98.74  & 98.83 & 98.65 & 99.48 & 98.15 \\
				\midrule
				\textbf{Ours}  & \textbf{98.12} & \textbf{99.01} & 98.83 & \textbf{99.2} & \textbf{99.79} & \textbf{98.81} \\
				\bottomrule
				\caption{ A quantitative comparison of the results obtained on the second version of proposed lung dataset test set using a variety of techniques for image segmentation. The best scores are highlighted with bold.
					\label{tbl_exp_imgsegmentation1} 
				}
			\end{tabular}
		}
	\end{center}
\end{table*}%

\subsection{Experimental Settings} \label{sec_Experiments_setting} \label{sec:sec_Experiments.Experimental Settings}
For both scenarios, we used a ResNet-50 encoder for all segmentation models. This choice was motivated by the balance between model complexity and performance, leveraging the pre-trained weights from ImageNet \cite{russakovsky2015imagenet} to facilitate transfer learning. The experiments were conducted on Google Colab, utilizing the Tesla T4 GPU. We employed the PyTorch framework for model training. The training batch size was set to 32, and we utilized the Adam optimizer with a learning rate of $10^{-4}$ for parameter optimization. The input images were resized to 320 $\times$ 320 pixels. 

\subsection{Evaluation Metrics} \label{sec_Experiments_metrics}
We employed several metrics to comprehensively evaluate the performance of our network. These metrics include measures for semantic segmentation tasks. Here, we provide details on the five metrics used for evaluation:
Intersection over Union (IoU) serves as the primary evaluation measure for semantic segmentation. Additionally, we calculate the following standard metrics \cite{fateh2021multilingual}:
\begin{align}
	& \text{Precision} = \frac{TP}{TP + FP} \\
	& \text{Recall} = \frac{TP}{TP + FN} \\
	& \text{F1-score} = \frac{2 \cdot TP}{2 \cdot TP + FP + FN} \\
	& \text{Accuracy} = \frac{TP + TN}{TP + TN + FP + FN} \\
	& \text{IoU} = \frac{TP}{TP + FP + FN} \\
	& \text{MCC} = \frac{ (TP \times TN) - (FP \times FN) }{ \sqrt{ (TP + FP)(TP + FN)(TN + FP)(TN + FN) } }
\end{align}

\begin{figure*}[!h]
	\begin{center}
		\begin{tabular}{M{3cm} M{2cm} M{2cm} M{2cm} M{2cm} } 
			&
			A &
			B &
			C &
			D \\ [0ex] 
			Input &
			\includegraphics[width=2.3cm,height=2.6cm]{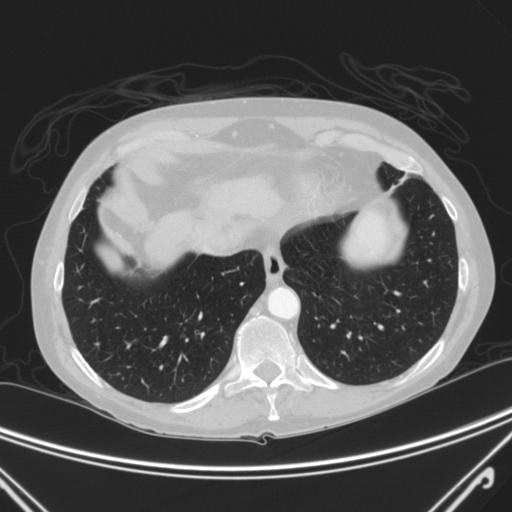} &
			\includegraphics[width=2.3cm,height=2.6cm]{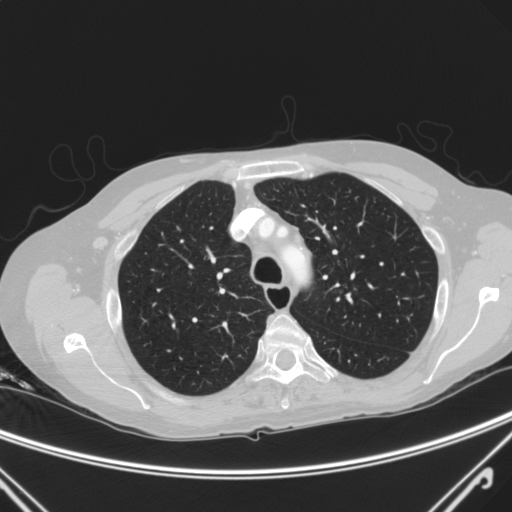} &
			\includegraphics[width=2.3cm,height=2.6cm]{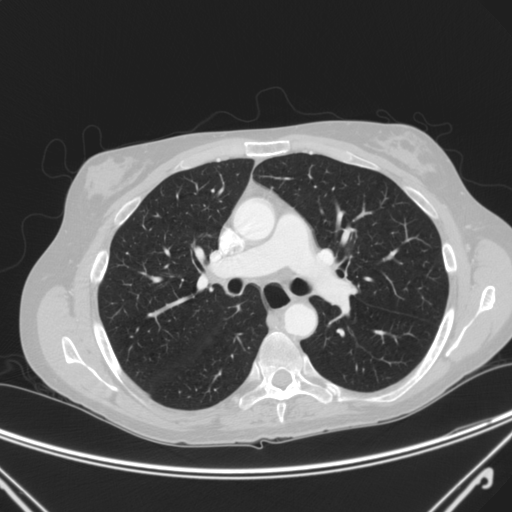} &
			\includegraphics[width=2.3cm,height=2.6cm]{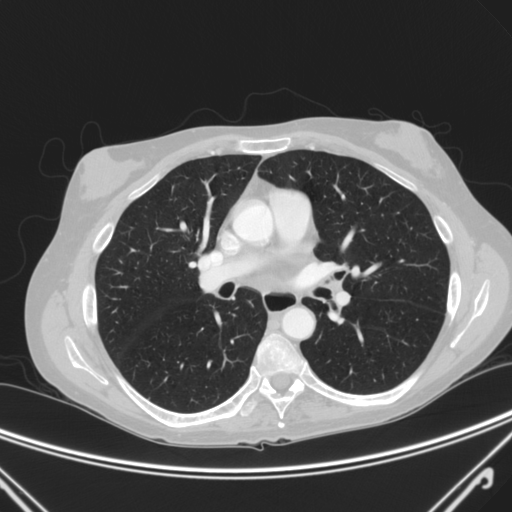} \\ [0ex] 
			Ground Truth &
			\includegraphics[width=2.3cm,height=2.6cm]{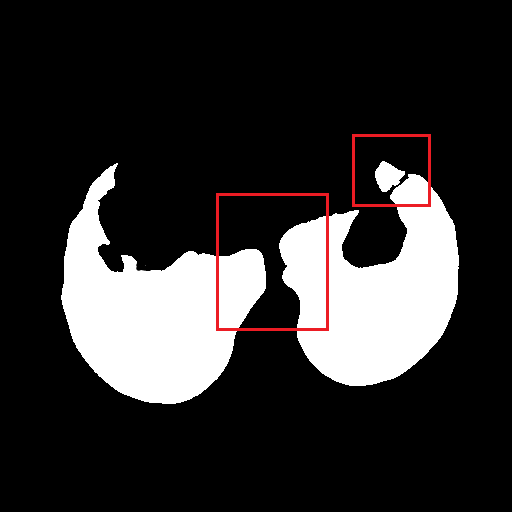} &
			\includegraphics[width=2.3cm,height=2.6cm]{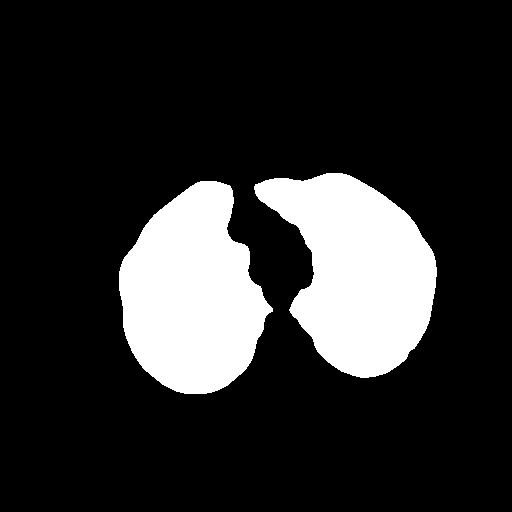} &
			\includegraphics[width=2.3cm,height=2.6cm]{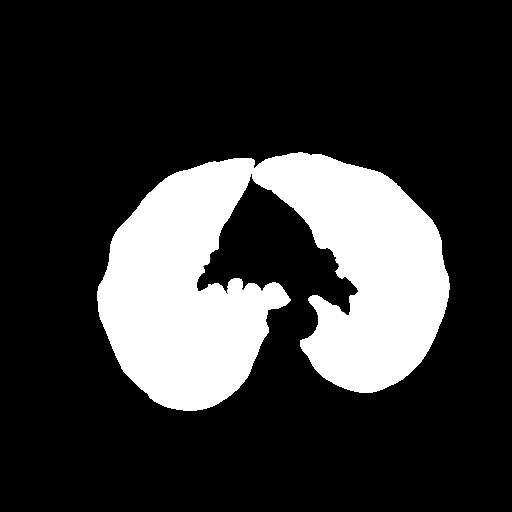} &
			\includegraphics[width=2.3cm,height=2.6cm]{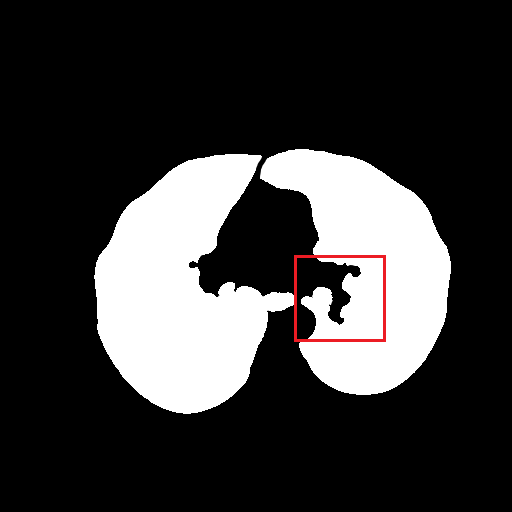} \\ [0ex] 
			Unet &
			\includegraphics[width=2.3cm,height=2.6cm]{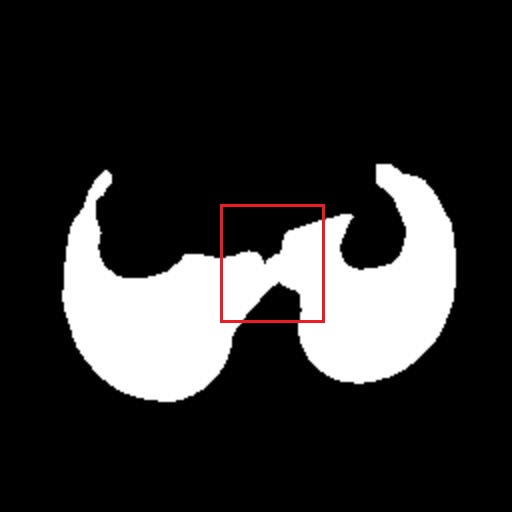} &
			\includegraphics[width=2.3cm,height=2.6cm]{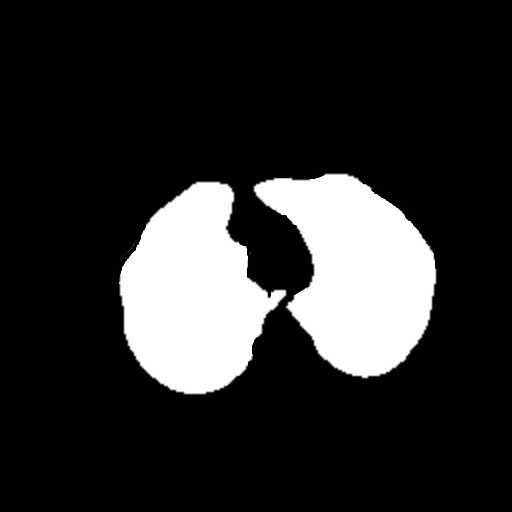} &
			\includegraphics[width=2.3cm,height=2.6cm]{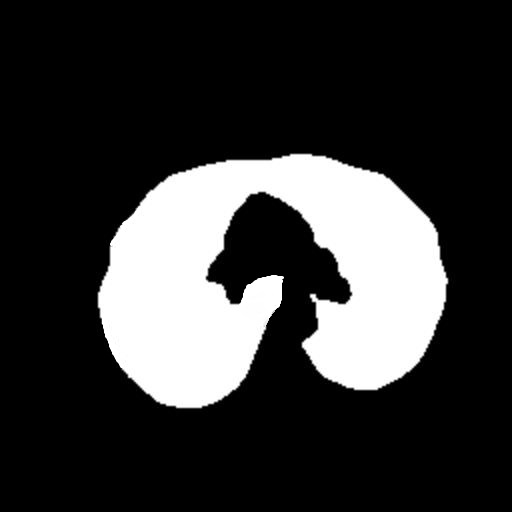} &
			\includegraphics[width=2.3cm,height=2.6cm]{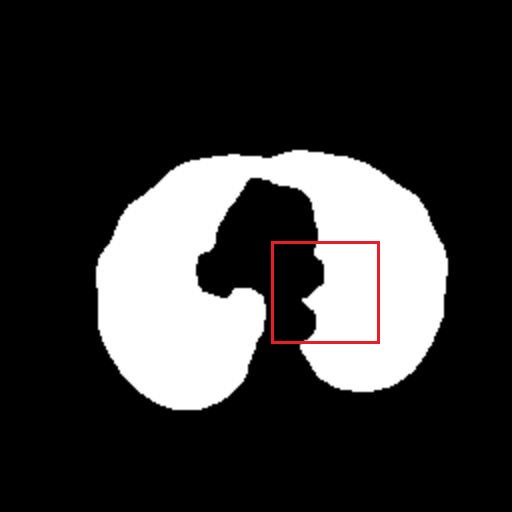} \\ [0ex] 
			NasNet &
			\includegraphics[width=2.3cm,height=2.6cm]{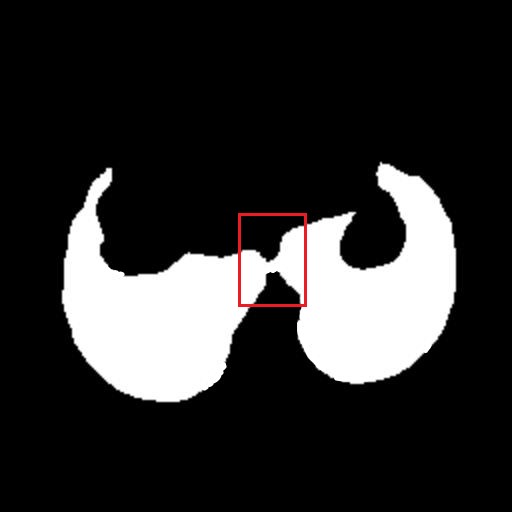} &
			\includegraphics[width=2.3cm,height=2.6cm]{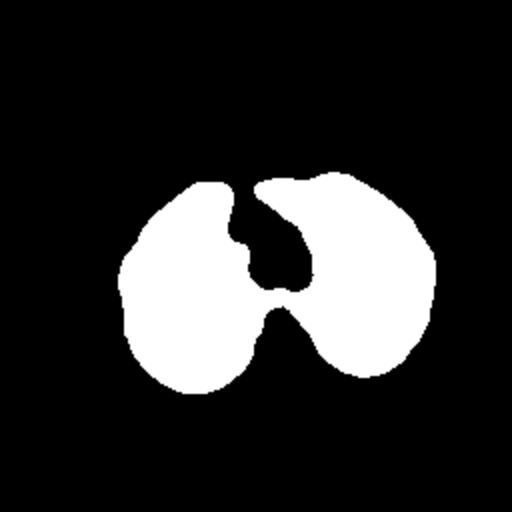} &
			\includegraphics[width=2.3cm,height=2.6cm]{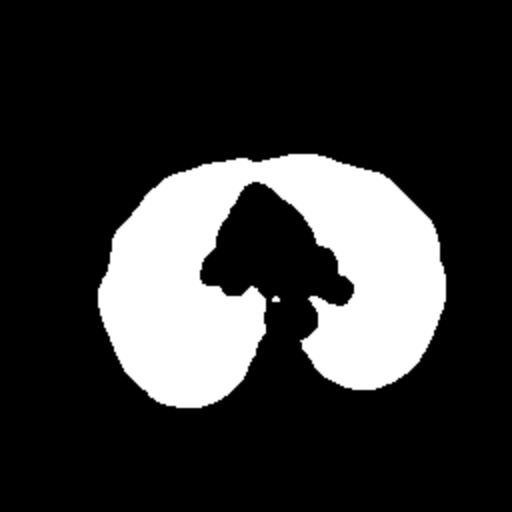} &
			\includegraphics[width=2.3cm,height=2.6cm]{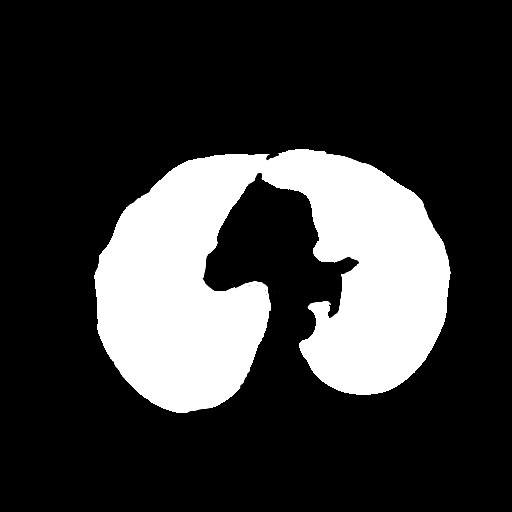} \\ [0ex]
			DABT-U-Net &
			\includegraphics[width=2.3cm,height=2.6cm]{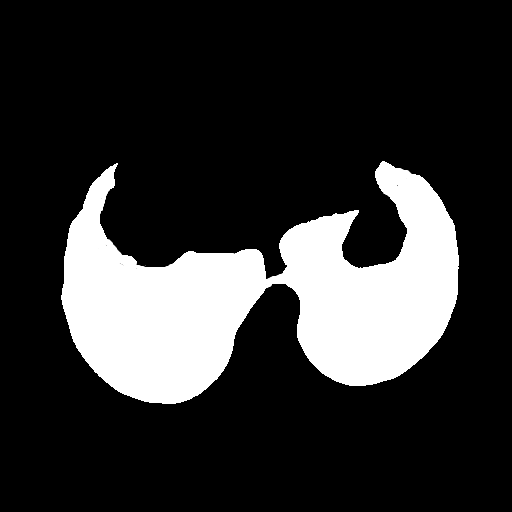} &
			\includegraphics[width=2.3cm,height=2.6cm]{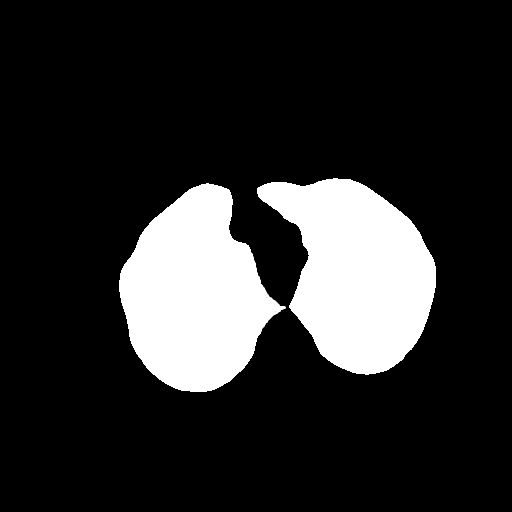} &
			\includegraphics[width=2.3cm,height=2.6cm]{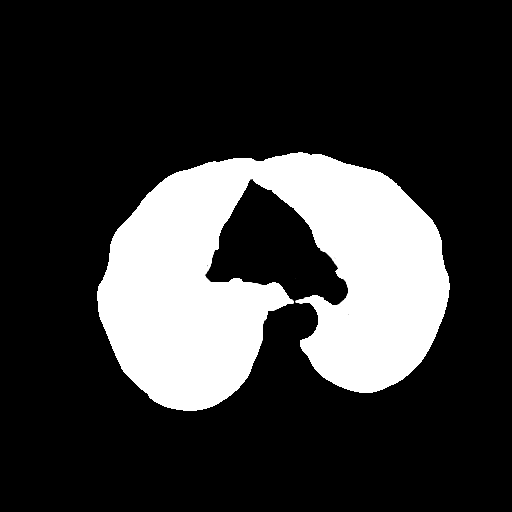} &
			\includegraphics[width=2.3cm,height=2.6cm]{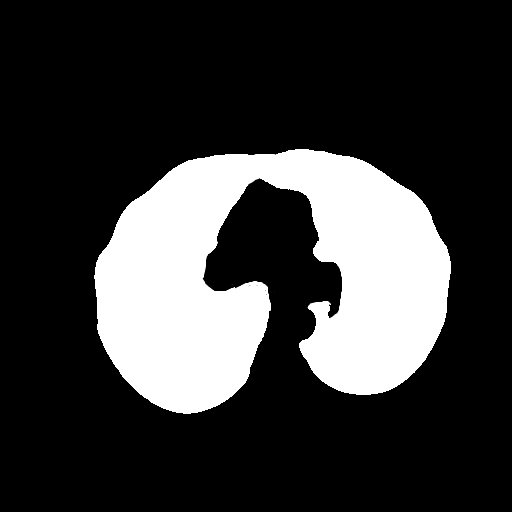} \\ [0ex]
			ABANet &
			\includegraphics[width=2.3cm,height=2.6cm]{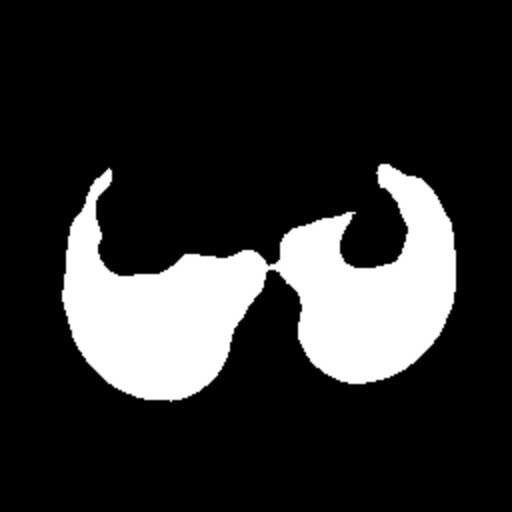} &
			\includegraphics[width=2.3cm,height=2.6cm]{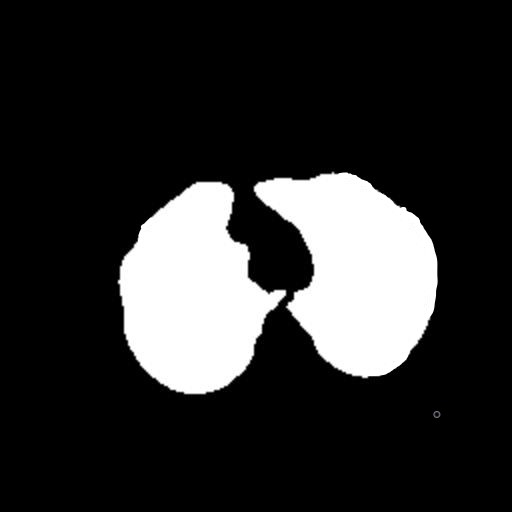} &
			\includegraphics[width=2.3cm,height=2.6cm]{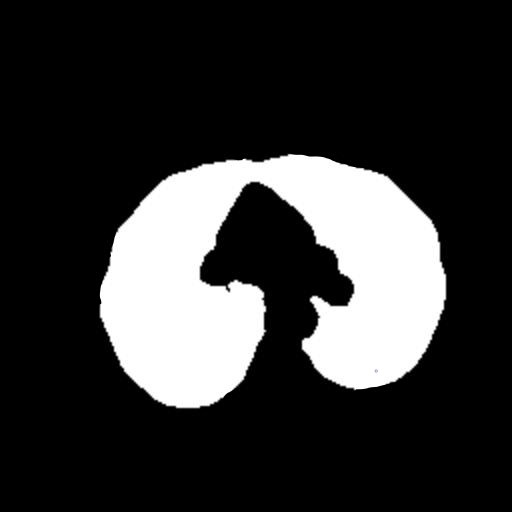} &
			\includegraphics[width=2.3cm,height=2.6cm]{./4_ex_aba.jpg} \\ [0ex]
			Ours &
			\includegraphics[width=2.3cm,height=2.6cm]{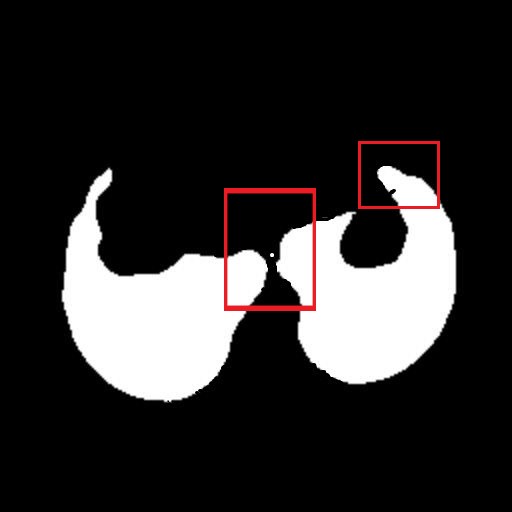} &
			\includegraphics[width=2.3cm,height=2.6cm]{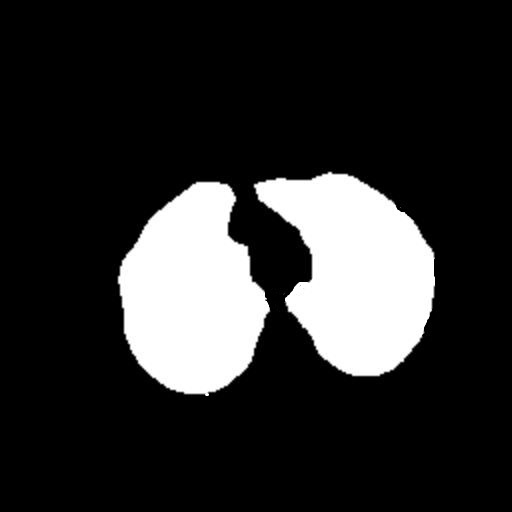} &
			\includegraphics[width=2.3cm,height=2.6cm]{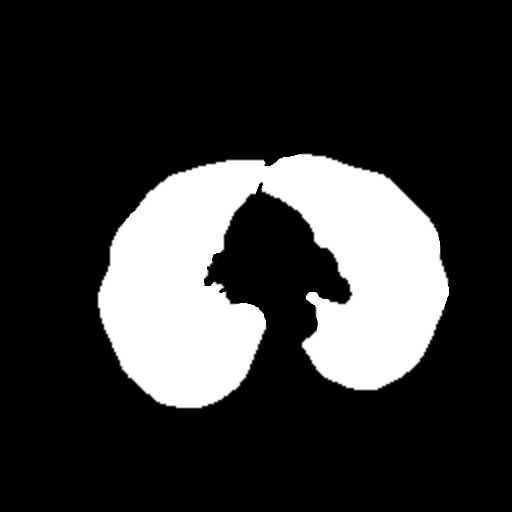} &
			\includegraphics[width=2.3cm,height=2.6cm]{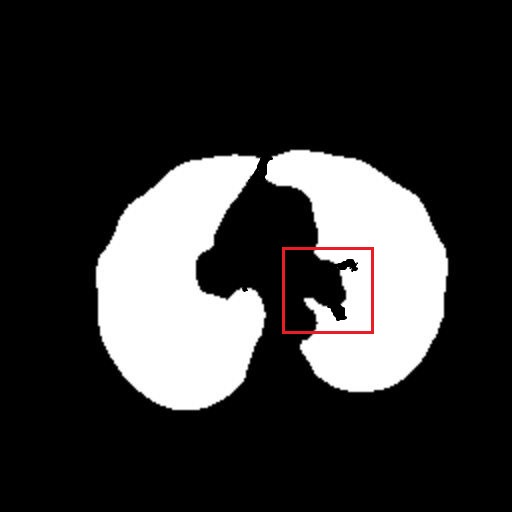} \\ [0ex]
		\end{tabular}
	\end{center}
	\caption{Qualitative comparison of lung dataset  results with cutting-edge image segmentation algorithms. The first two rows display the original photos and the matching ground truth. Rows 3 to 8 illustrate the segmentation results respectively derived from the proposed method, Unet, RU-Net, ResNet34-Unet, BCDU-Net, and ResBCDUnet.\label{fig_exp_resultimg}}
\end{figure*}

\subsection{Comparison with state-of-the-art models}\label{sec_comparison}
On our lung segmentation dataset, we conduct a comparison of our network with other state-of-the-art approaches using the same training configuration of 1600 training images. Segmentation models are Unet\cite{ronneberger2015u}, RU-Net\cite{alom2018nuclei}, ResNet34-Unet\cite{lau2020automated},  BCDU-Net\cite{azad2019bi}, ResBCDUnet\cite{jalali2021resbcdu}, ABANet\cite{rezvaniabanet}, NasNet \cite{zhang2022automatic} and DABT-U-Net \cite{jalali2024dabt}.

\textbf{Quantitative Evaluation:} 
as shown in Table \ref{tbl_exp_imgsegmentation}, the evaluation metrics were calculated and summarized for quantitative comparisons on first version of proposed dataset.
As can be observed from Table \ref{tbl_exp_imgsegmentation}, our proposed FusionLungNet achieved 98.04, 99.02, 98.43, 99.73 and 98.67 in terms of IoU, F1-score, recall, accuracy and MCC, respectively, which were superior to those of other methods. The IoU and F1-Score of the proposed network were approximately 3\% higher than those of the classical networks like Unet, RU-Net, and ResNet34-Unet, which implied that the hybrid loss can help efficiently. Compared to newer methods like ABANet and DABT-U-Net, the proposed network showed marginal improvements in the result, which indicates that  various proposed modules are suitable for medical image segmentation.

To further validate the effectiveness of our approach, we conducted additional experiments using the first and second version of our dataset, which comprises 2,350 labeled images with applied preprocessing techniques as detailed in Section \ref{sec_preprocessing}. The results of these experiments are presented in Table \ref{tbl_exp_imgsegmentation1}.

As illustrated in Table~\ref{tbl_exp_imgsegmentation1}, our FusionLungNet model consistently outperforms the baseline models across all evaluation metrics when trained on the enhanced dataset with preprocessing. Specifically, FusionLungNet achieved an IoU of 98.12\%, F1-score of 99.01\%, Precision of 98.83\%, Recall of 99.2\%, Accuracy of 99.79\%, and MCC of 98.81\%. These improvements underscore the benefits of increasing the dataset size and applying preprocessing techniques, which collectively contribute to more robust and accurate lung segmentation performance.

Comparing the results from Version 1 (Table~\ref{tbl_exp_imgsegmentation}) and Version 2 (Table~\ref{tbl_exp_imgsegmentation1}), it is evident that the expanded dataset and preprocessing steps lead to significant enhancements in model performance. The higher Recall and Accuracy values indicate better detection capabilities and overall correctness, while the increased IoU and MCC reflect improved segmentation overlap and correlation with ground truth labels.

\begin{figure*}[h!]
	\begin{center}
		\begin{tabular}{M{3cm} M{2cm} M{2cm} M{2cm} M{2cm} } 
			  &
			A &
			B &
			C &
			D \\ [0ex] 
			Input &
			\includegraphics[width=2.3cm,height=2.6cm]{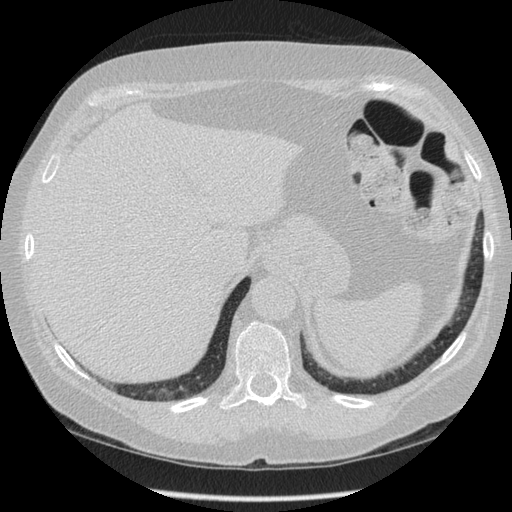} &
			\includegraphics[width=2.3cm,height=2.6cm]{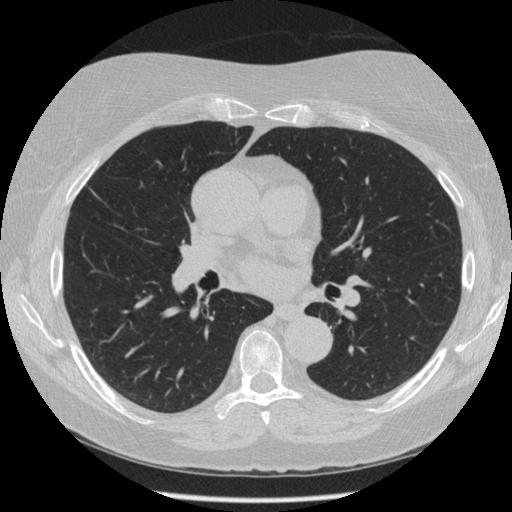} &
			\includegraphics[width=2.3cm,height=2.6cm]{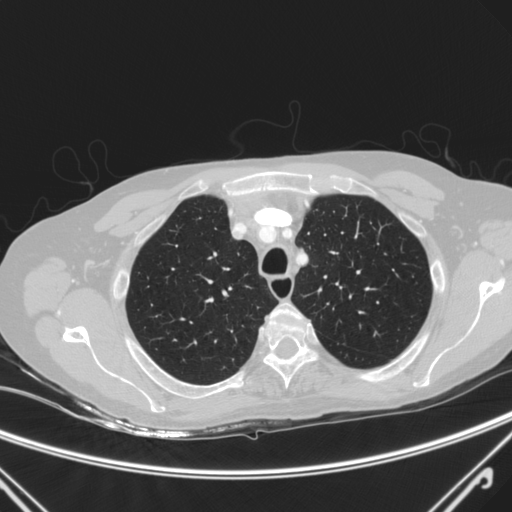} &
			\includegraphics[width=2.3cm,height=2.6cm]{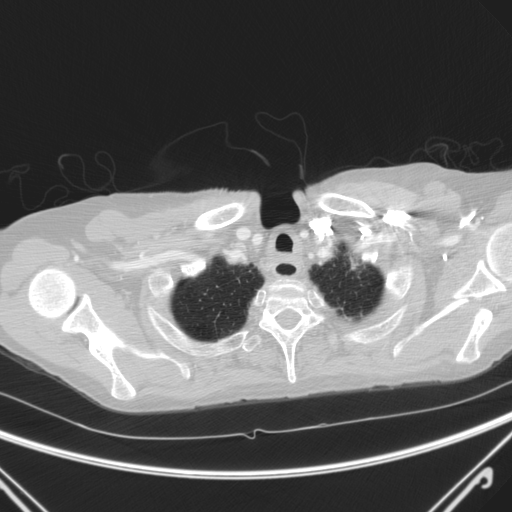} \\ [0ex] 
			Ground Truth &
			\includegraphics[width=2.3cm,height=2.6cm]{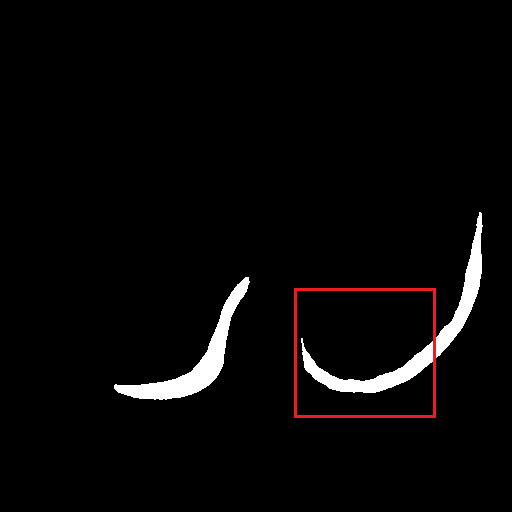} &
			\includegraphics[width=2.3cm,height=2.6cm]{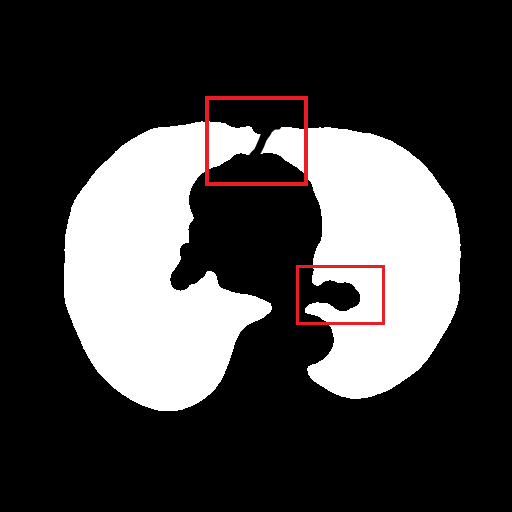} &
			\includegraphics[width=2.3cm,height=2.6cm]{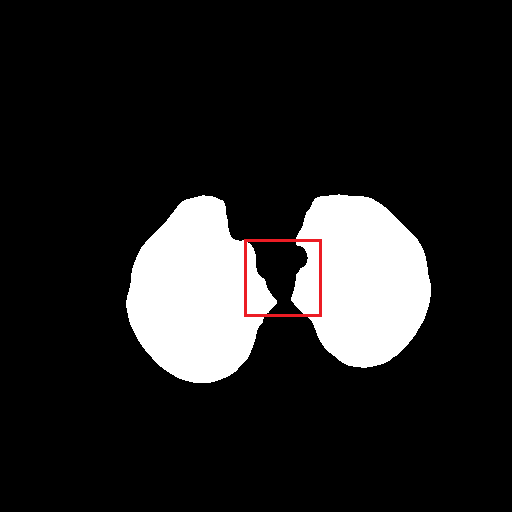} &
			\includegraphics[width=2.3cm,height=2.6cm]{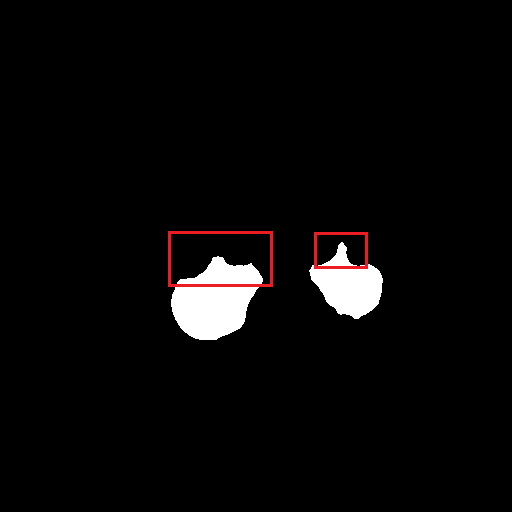} \\ [0ex] 
			base &
			\includegraphics[width=2.3cm,height=2.6cm]{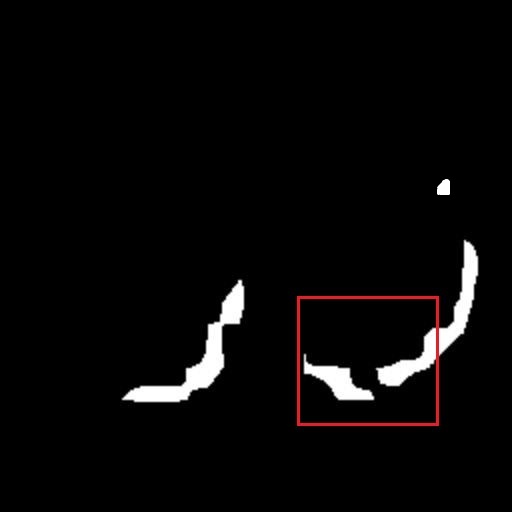} &
			\includegraphics[width=2.3cm,height=2.6cm]{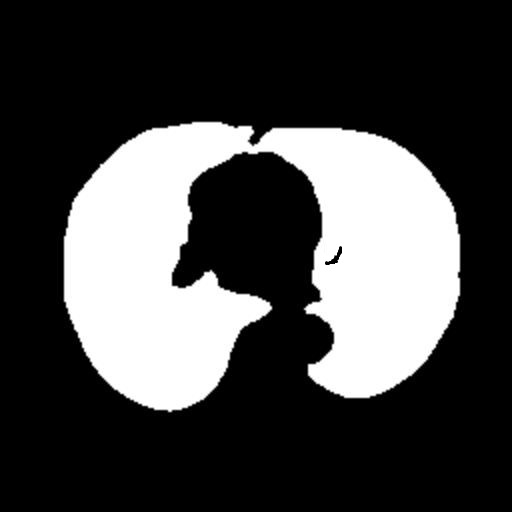} &
			\includegraphics[width=2.3cm,height=2.6cm]{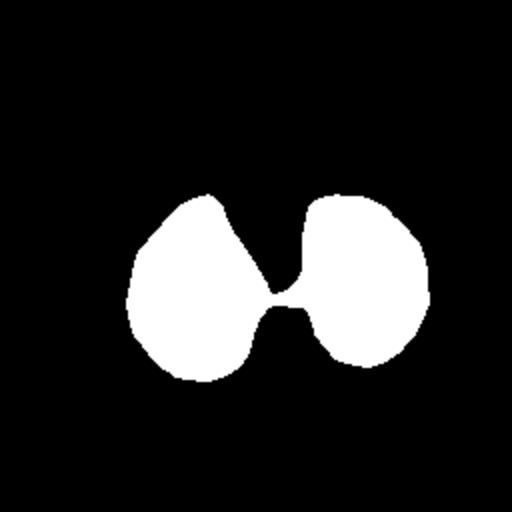} &
			\includegraphics[width=2.3cm,height=2.6cm]{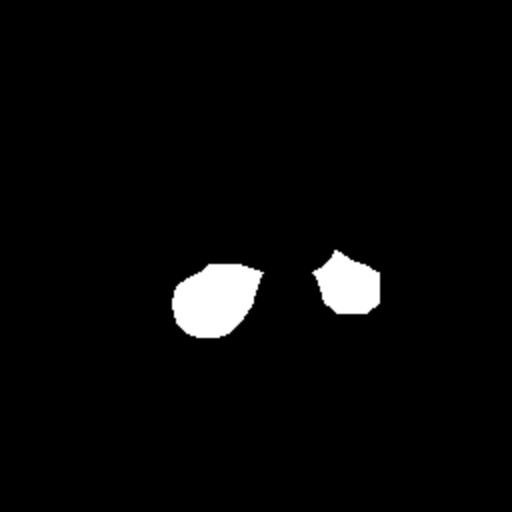} \\ [0ex]
			base + residual refinement module &
			\includegraphics[width=2.3cm,height=2.6cm]{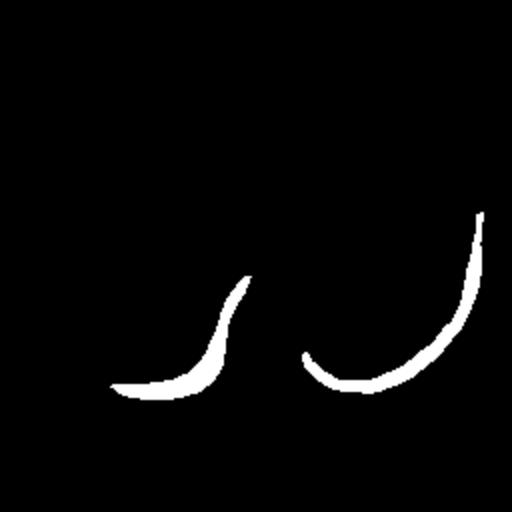} &
			\includegraphics[width=2.3cm,height=2.6cm]{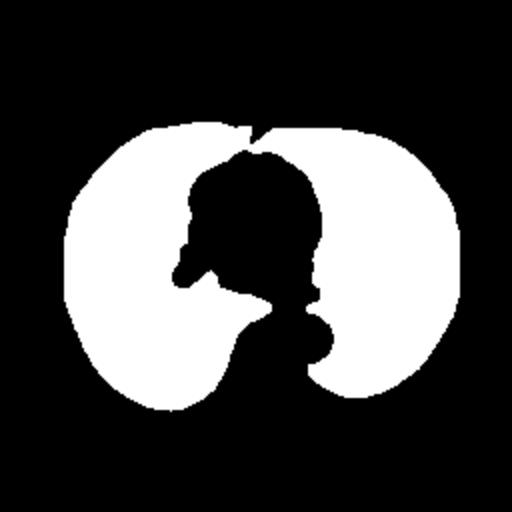} &
			\includegraphics[width=2.3cm,height=2.6cm]{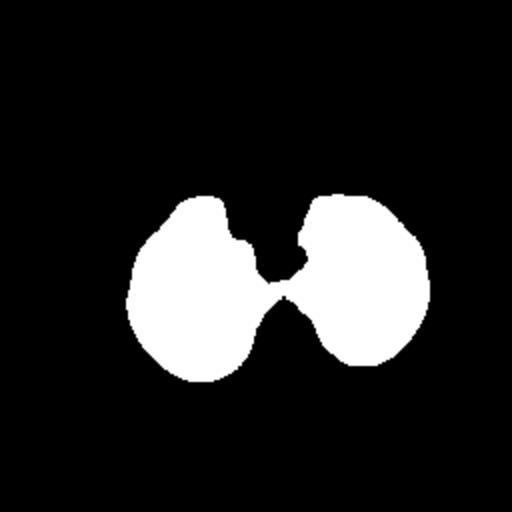} &
			\includegraphics[width=2.3cm,height=2.6cm]{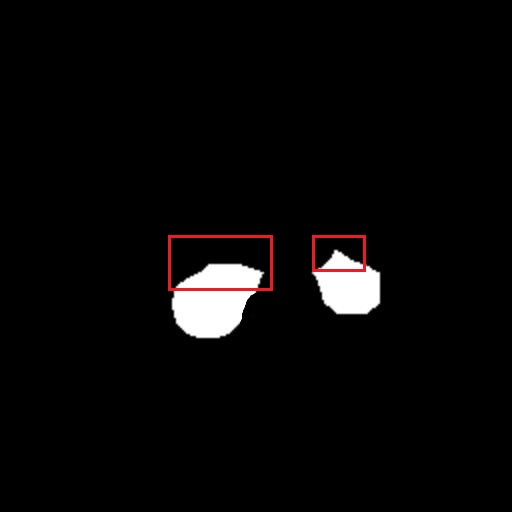} \\ [0ex]
			base + MFF &
			\includegraphics[width=2.3cm,height=2.6cm]{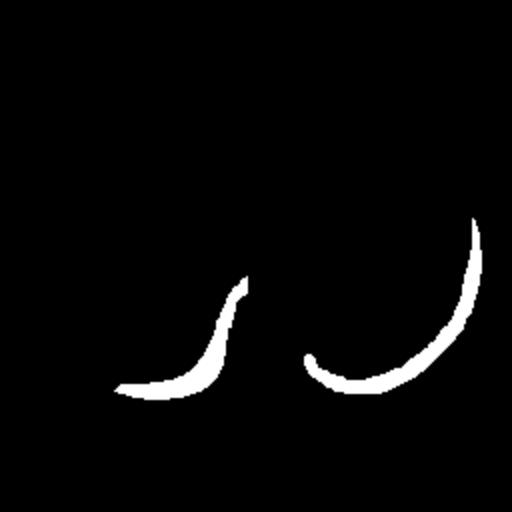} &
			\includegraphics[width=2.3cm,height=2.6cm]{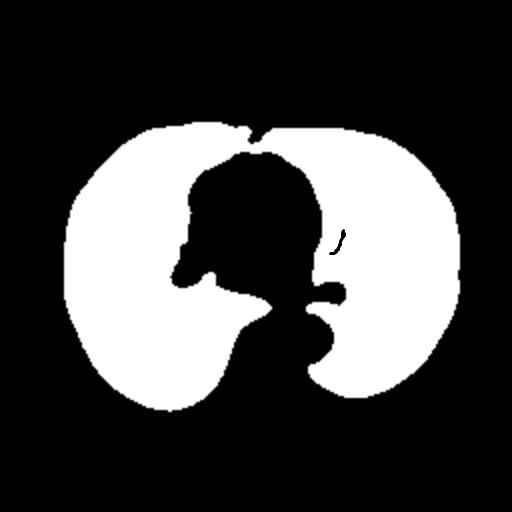} &
			\includegraphics[width=2.3cm,height=2.6cm]{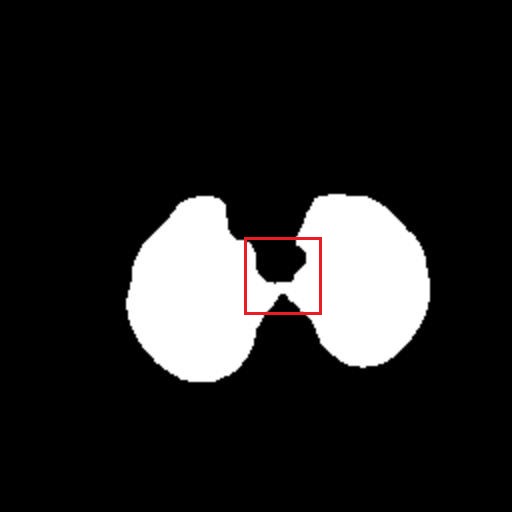} &
			\includegraphics[width=2.3cm,height=2.6cm]{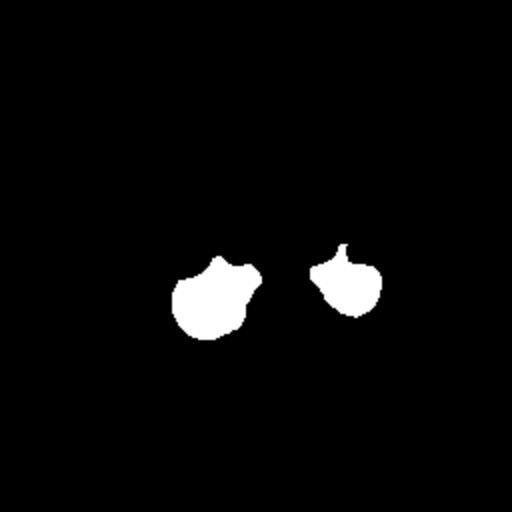} \\ [0ex]
			base + MFF + self refinement block &
			\includegraphics[width=2.3cm,height=2.6cm]{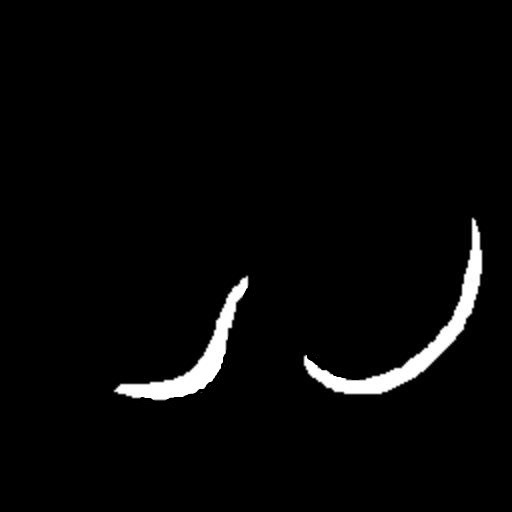} &
			\includegraphics[width=2.3cm,height=2.6cm]{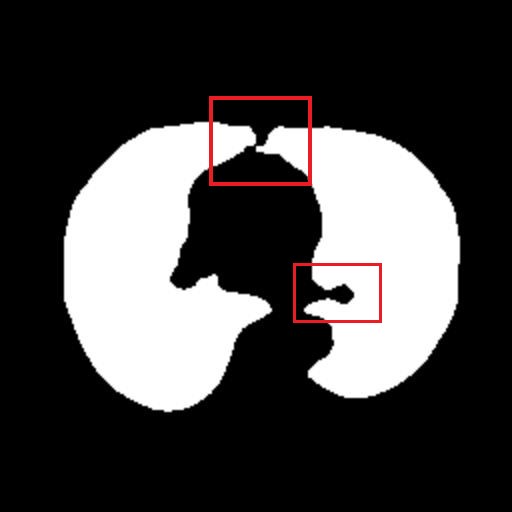} &
			\includegraphics[width=2.3cm,height=2.6cm]{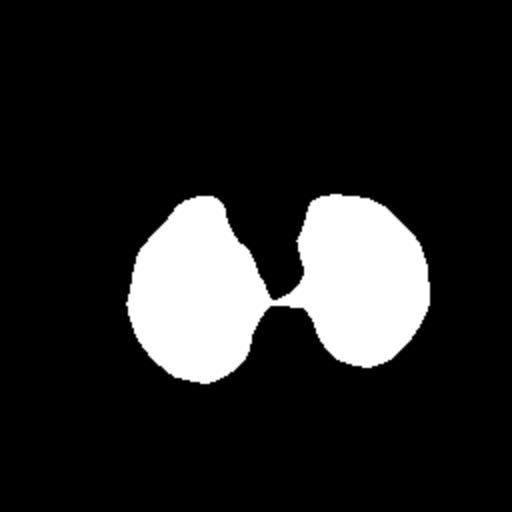} &
			\includegraphics[width=2.3cm,height=2.6cm]{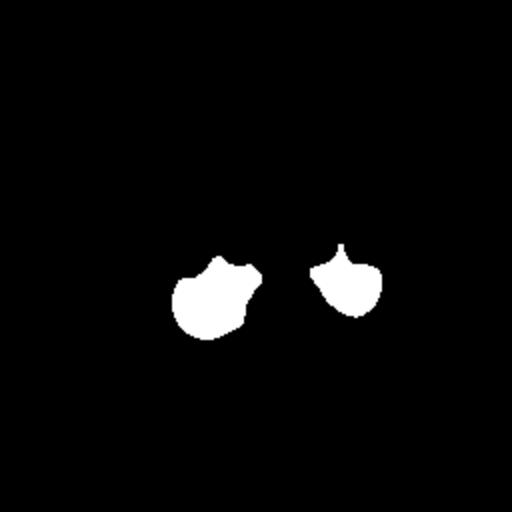} \\ [0ex]
			base+MFF+ self refinement block + residual refinement module &
			\includegraphics[width=2.3cm,height=2.6cm]{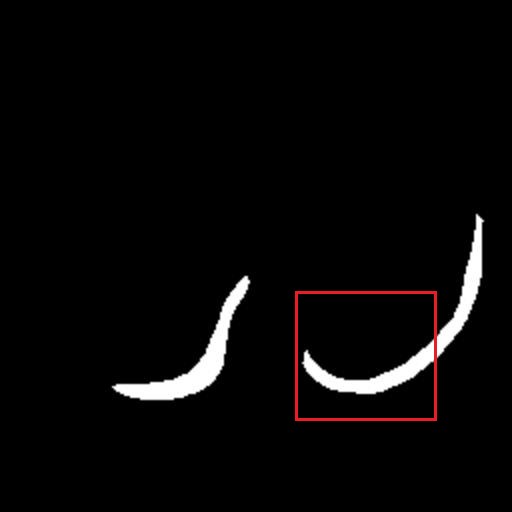} &
			\includegraphics[width=2.3cm,height=2.6cm]{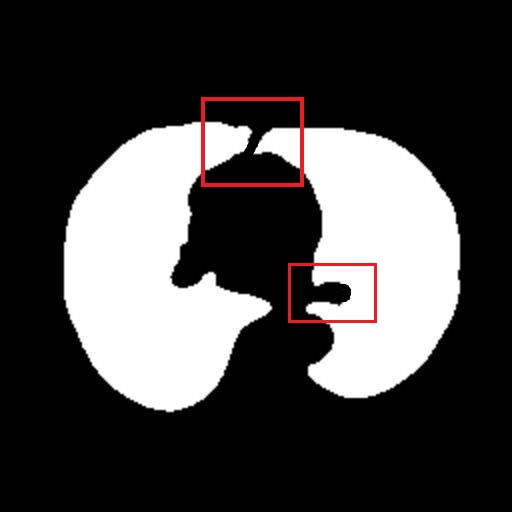} &
			\includegraphics[width=2.3cm,height=2.6cm]{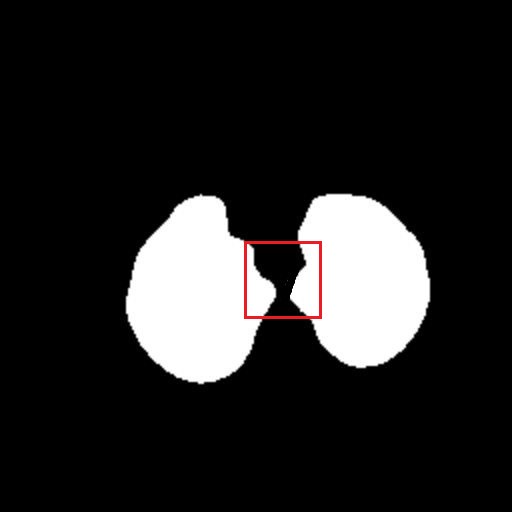} &
			\includegraphics[width=2.3cm,height=2.6cm]{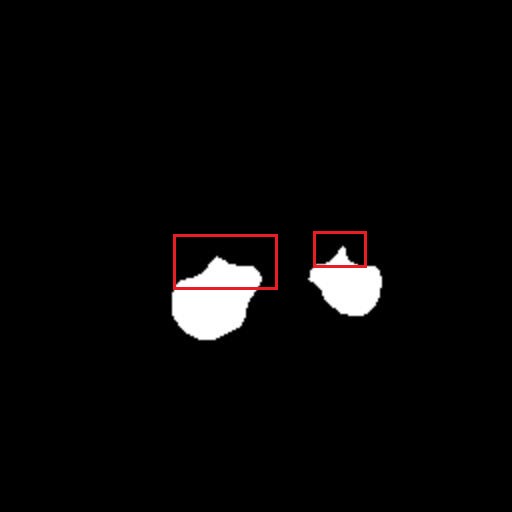} \\ [0ex]
		\end{tabular}
	\end{center}
	\caption{Qualitative comparison of lung dataset  results with cutting-edge image segmentation algorithms. The first two rows display the original photos and the matching ground truth. Rows 3 to 8 illustrate the segmentation results respectively derived from the proposed method, Unet, RU-Net, ResNet34-Unet, BCDU-Net, and ResBCDUnet.\label{fig_exp_imgsegmentation}}
\end{figure*}

\textbf{Qualitative Evaluation:}
In Figure \ref{fig_exp_resultimg}, we present a comparison of experiments conducted on our proposed dataset, showcasing typical samples from the test set. Among the evaluated networks Unet, NasNet, DABT-U-Net, and ABANet, our network exhibits strong performance in predicting the fuzzy edges seen on the right side of sample D in figure \ref{fig_exp_resultimg}, closely resembling the ground truth. However, DABT-U-Net overly emphasizes boundary information, neglecting to predict the overall area of the sample, which impacts its accuracy. Conversely, Unet and NasNet incorrectly identify the background as lung tissue in the small region on the center  of sample A in figure \ref{fig_exp_resultimg}. In contrast, our approach demonstrates minimal deviation from the ground truth, highlighting its superior performance in handling challenging cases.

Overall, while state-of-the-art methods often suffer from over-segmentation or under-segmentation when dealing with complex scenes, our approach excels at precisely preserving lung boundaries and structures. Notably, our network proves more robust in describing boundaries and edges compared to existing approaches.



\begin{table*}[!htbp]
	\begin{center}
		\caption{ The performance of different configurations of FusionLungNet. The best scores are highlighted with bold.
			\label{tbl_exp_config}}
		
		\begin{tabular}{p{9cm}ccccc}
			\toprule
			Method & IoU  & F1-score & Precision & Recall & Acc\\
			\midrule
			baseline    & 95.57   & 96.67  & 97.54  & 95.82 & 97.15  \\
			baseline + MFF    & 96.64   & 97.03  & 96.94 & 97.14 & 98.1  \\
			baseline + residual refinement module    & 95.91   & 96.88  & 97.28  & 96.37 & 97.256  \\
			baseline +  MFF + self refinement block    & 97.18   & 98.54  & 98.3 & 98.83 & 99.52  \\
			baseline + MFF + self refinement block + \\ residual refinement module    & \textbf{98.04 }  & \textbf{99.02}  & \textbf{98.65} & \textbf{99.41} & \textbf{99.73}   \\
			\bottomrule
		\end{tabular}
	\end{center}
\end{table*}%

\begin{table*}[!htbp]
	\begin{center}
		\resizebox{\columnwidth}{!}{
			\begin{tabular}{lccc}
				\toprule
				&  & Input size & \\
				\cmidrule{2-4}
				& 160 × 160 & 320 × 320 & 640 × 640  \\
				\specialrule{1pt}{1pt}{1pt}
				IoU      & 97.93   & \textbf{98.04}  & 97.81     \\
				\bottomrule
				\caption{ FusionLungNet performance with various input picture resolutions.
					\label{tbl_exp_inputsize}}
				
		\end{tabular}}
	\end{center}
\end{table*}%
\begin{table*}[!htbp]
	\begin{center}
		\caption{  The performance of different optimizer.
			\label{tbl_exp_optimizer}}
		
		\begin{tabular}{lcccc}
			\toprule
			Optimizer & IoU  & F1-score & Acc\\
			\midrule
			SGD    & 96.8   & 97.58  & 98.65    \\
			Adamax     & 97.5   & 98.71  & 99.77   \\
			RMSprop    & 97.63   & 99.05  & \textbf{99.79}    \\
			Adam     & \textbf{98.04}   & \textbf{99.02}  & 99.73    \\
			\bottomrule
		\end{tabular}
	\end{center}	
\end{table*}%
\begin{table*}[!htbp]
	\begin{center}
		\captionsetup{justification=justified}
		\caption{ Effect of various loss functions on FusionLungNet performance.
			\label{tbl_exp_loss}}
		
		\begin{tabular}{lcccc}
			\toprule
			Losses & IoU  & F1-score &  Acc\\
			\midrule
			Focal    & 95.42   & 95.18  & 96.4    \\
			IoU     & 95.9   & 95.81  & 96.64   \\
			Focal + IoU    & 96.27   & 97.5  & 97.74    \\
			Focal + SSIM + IoU     & \textbf{98.04}   & \textbf{99.02}  & \textbf{99.73}      \\
			\bottomrule
		\end{tabular}
	\end{center}	
\end{table*}%
\begin{table*}[!htbp]
	\begin{center}
		\caption{  effectiveness of preprocessing stage.
			\label{tbl_exp_preproc}}
		
		\begin{tabular}{lccccc}
			\toprule
			Model & Preprocessing & IoU  & F1-score & Acc\\
			\midrule
			FusionLungNet & \checkmark    & \textbf{98.12}   & 99.01  & 99.79    \\
			FusionLungNet &  $\times$   & 98.07   & \textbf{99.02}  &\textbf{ 99.75}   \\
			\bottomrule
		\end{tabular}
	\end{center}	
\end{table*}%
\subsection{Ablation study}\label{sec_ablation}
In this section, to further evaluate the effectiveness of FusionLungNet, we conducted ablation studies using the proposed dataset as examples.

To verify the effectiveness of each component, we take the encoder with the CAA module and the decoder as the baseline and compare its results with four modalities of our lung segmentation: Baseline + MFF, baseline + residual refinement module, baseline + MFF + self refinement module, and baseline + MFF + self refinement module + residual refinement module. The implicit parameters of all models, such as batch size and learning rate, are set the same to ensure the fairness of the experiment. The experiment is carried out on our lung segmentation dataset. The F1-score, IoU, precision, recall, and accuracy are selected as evaluation metrics.

The experimental results are shown in Table \ref{tbl_exp_config}. We first introduce the MFF block into the baseline. Compared with the baseline, the introduction of the MFF block improves the IoU, F1-score, and accuracy by 0.7\%, 0.36\%, and 1\%, respectively. Other components improved the baseline to a certain extent. The scores of the baseline with all components in the five evaluation metrics are higher than those of other network models. Compared with the baseline, the segmentation performance of the proposed method is improved by 3.4\% in IoU, 2.65\% in F1-score, 1.13\% in precision, 3.7\% in recall, and 2.6\% in accuracy.

Figure \ref{fig_exp_imgsegmentation} shows a graphical visualization of the results of the ablation experiment, displaying four examples of images and the segmentation masks corresponding to the different component modules. From the segmentation results of the first test instance, it can be seen that there are varying degrees of under-segmentation compared to the base version of our entire model, while the proposed network model had better feature representation. It is evident from the segmentation results shown in Figure \ref{fig_exp_imgsegmentation} that our entire framework provided clearer segmentation of the boundary information than either a single or a combination of any component.

To evaluate the stability and performance of the network, we trained FusionLungNet with three different input sizes:  160 $\times$ 160,  320 $\times$ 320 and 640 $\times$ 640 pixels. The selection of these various input dimensions was aimed at assessing the model's performance under different conditions that reflect practical scenarios in medical imaging. Specifically, smaller input sizes, such as 160 × 160 pixels, allow for faster processing and reduced computational demands. Conversely, larger input sizes, like 640 × 640 pixels, capture more detailed features, which may enhance segmentation accuracy but require more computational resources. Testing these input sizes helps ensure the model's robustness and generalizability across varying image qualities and resolutions.

Based on the data presented in Table \ref{tbl_exp_inputsize}, the input size of 320 $\times$ 320 pixels yielded the best IoU score of 98.04\%, indicating an optimal balance between computational efficiency and segmentation accuracy.

In this article, we use Adam as our optimizer. We also comprise different major optimizers: Adam, Adamax, RMSprop, and SGD in Table \ref{tbl_exp_optimizer}. the Adam optimizer outperformed other optimizers.

In order to illustrate the efficacy of our suggested hybrid loss, we perform an ablation study on losses with the same experimental setup. Table \ref{tbl_exp_loss} provides the comparing results. It can be observed that both focal loss and IoU loss Have almost similar performance in terms of F1-score and accuracy metrics. However, when it comes to the IoU measure, IoU loss outperforms focal loss. When we mix focal loss and IoU loss, all metrics increased slightly. To propose a final hybrid loss, we use SSIM with two other losses. Table \ref{tbl_exp_loss}  indicates that by equipping hybrid loss on FusionLungNet performance improve greatly. It is due to utilizing SSIM loss and obtaining more detailed boundary prediction. 

To evaluate the impact of the preprocessing stage on the performance of our proposed model, FusionLungNet, an ablation study was conducted. From the results in Table \ref{tbl_exp_preproc}, it can be observed that the inclusion of preprocessing steps led to a slight improvement in IoU and Accuracy. The ablation study confirms that the preprocessing step, although not dramatically altering performance, adds a beneficial enhancement to the overall segmentation accuracy, supporting its inclusion in the pipeline.

\section{Conclusion} \label{sec_conclusion}
This paper introduces a novel framework FusionLungNet for lung segmentation. The proposed approach is based on an encoder-decoder architecture that utilizes various modules for the segmentation process in lung CT images, incorporating an innovative hybrid loss. To effectively mitigate the defects associated with the direct fusion of different layers and ensure the generation of more discriminative features, we propose a self refinement module. Additionally, we introduce the MFF (Multi-Feature Fusion) block designed to integrate features from various levels. Finally, for further refinement of the reconstructed image, we propose the residual refinement module. Our experiments on the proposed datasets demonstrate that FusionLungNet outperforms eight state-of-the-art methods across different evaluation metrics.

Advantages of FusionLungNet include its ability to capture detailed features at multiple scales, enhancing segmentation accuracy while maintaining computational efficiency. The incorporation of a hybrid loss function, which combines Focal Loss, Structural Similarity Index (SSIM), and IoU Loss, contributes significantly to the model's robustness in handling class imbalance and improving boundary delineation.

While the FusionLungNet framework has shown promising results, there are several limitations to our approach. First, the model's performance might be constrained by the class imbalance present in the dataset, where lung regions constitute a small proportion of the entire image. This imbalance can lead to less accurate predictions for minority classes. Second, although our method integrates multiple innovative modules, the increased complexity can result in higher computational costs and longer training times. Finally, the proposed method's effectiveness may vary with different datasets, and its generalizability needs further validation across a wider range of clinical scenarios and imaging modalities.

For future work, several directions can be pursued to address these limitations and enhance the FusionLungNet framework. One potential improvement is to implement advanced data augmentation techniques to alleviate the class imbalance issue. Additionally, exploring more efficient network architectures could help reduce computational costs while maintaining high performance. Incorporating domain adaptation techniques might also improve the model's generalizability to diverse datasets.

\section*{CRediT author statement}
\textbf{S. Rezvani}: Conceptualization, Methodology, Software, Writing – Original Draft, Formal Analysis, Visualization, resources. \textbf{M. Fateh}: Validation, Conceptualization, Supervision, Review \& Editing, Methodology, Investigation, Project administration. \textbf{Y. Jalali}: Data Curation, Writing – Original Draft, Formal Analysis, resources. \textbf{A. Fateh}: Supervision, Writing – Review \& Editing, Validation, Visualization, Investigation.

\clearpage
\bibliographystyle{IEEEtran}
\bibliography{references}
\end{document}